\documentclass[%
 reprint,
 amsmath,amssymb,
 aps,
]{revtex4-2}

\usepackage{graphicx}
\usepackage{dcolumn}
\usepackage{amsmath}
\usepackage{mathtools}
\usepackage{amssymb}
\usepackage{hyperref}
\usepackage{braket}
\usepackage{mathtools}
\usepackage{xcolor}
\usepackage{bm}

\begin{document}


\title{Geometric measure of quantum complexity in cosmological systems}

\author{Satyaki Chowdhury\textsuperscript{1}}
\email{satyaki.chowdhury@doctoral.uj.edu.pl}
\author{Martin Bojowald\textsuperscript{2}}
\email{bojowald@psu.edu}
\author{Jakub Mielczarek\textsuperscript{1}}
\email{jakub.mielczarek@uj.edu.pl}

\affiliation{\textsuperscript{1}Institute of Theoretical Physics, Jagiellonian University, Lojasiewicza 11, 30-348 Cracow, Poland}
\affiliation{\textsuperscript{2}Doctoral School of Exact and Natual Sciences, Jagiellonian University, Lojasiewicza 11, 30-348 Cracow, Poland}
\affiliation{\textsuperscript{3}Institute for Gravitation and the Cosmos, The Pennsylvania State University,
104 Davey Lab, University Park, PA 16802, USA}

\begin{abstract}
In Nielsen's geometric approach to quantum complexity, the introduction of a suitable 
geometrical space, based on the Lie group formed by fundamental operators, facilitates 
the identification of complexity through geodesic distance in the group manifold. Earlier 
work has shown that the computation of geodesic distance can be challenging for Lie groups 
relevant to harmonic oscillators.  Here, this problem is approached by working to leading 
order in an expansion by the structure constants of the Lie group. An explicit formula 
for an upper bound on the quantum complexity of a harmonic oscillator Hamiltonian with 
time-dependent frequency is derived. Applied to a massless test scalar field on a 
cosmological de Sitter background, the upper bound on complexity as a function of 
the scale factor exhibits a logarithmic increase on super-Hubble scales. This result 
aligns with the gate complexity and earlier studies of de Sitter complexity. 
It demonstrates the consistent application of Nielsen complexity to 
quantum fields in cosmological backgrounds and paves the way for further applications.
\end{abstract}

\maketitle

\section{Introduction} 

The question of whether a computation is easy or hard is profoundly relevant across 
various contexts, particularly when considering physical processes from a computational 
perspective. The complexity of dynamical quantum processes, interpreted as quantum computation, 
stands as a focal point in quantum information theory. As implied by its name, the complexity 
of quantum computation can be conceptualized as the challenge inherent in constructing a 
quantum algorithm capable of executing a desired task. Technically, the complexity of quantum 
computations is quantified by the minimal number of elementary operations (quantum gates) 
needed sequentially to accomplish the intended computation. However, extending this notion 
to physical systems is nontrivial and beset with significant challenges. Besides contending 
with infinite-dimensional Hilbert spaces, numerous decisions must be made. For example, 
when evaluating the complexity of unitary operators, one must determine the level of precision 
required for their approximation using operators from a predefined set of gates, necessitating 
the introduction of a ``tolerance" term.

The notion of complexity that we find the most attractive and will apply to study quantum dynamics 
is the one proposed by Nielsen in \cite{Nielsen_2006,https://doi.org/10.48550/arxiv.quant-ph/0502070,https://doi.org/10.48550/arxiv.quant-ph/0701004}.  Nielsen \textit{et al.\ }provided a geometrical 
interpretation of quantum complexity. They identified the length of the minimal geodesic in a certain 
curved geometry with complexity, which is defined for any unitary operator as the length of the shortest 
geodesic on the group of unitaries that connects the identity to the operator in question. To 
determine the geodesics, one typically introduces ``cost factors'' in the metric that scale up the contributions to the 
line element coming from the ``hard'' generators as opposed to those from the ``easy'' generators. 
Physically, the notion of \textit{hard} or \textit{easy} may come from the difficulty associated with 
implementing non-local operations rather than local ones. 

The significance of quantum complexity in the high-energy physics community became apparent in work by
Susskind \emph{et al.} on searching for suitable observables to probe the physics behind the horizon 
of a black hole. The crucial motivation that led to the search was that the entanglement entropy 
saturates as the black hole thermalizes \cite{Hartman:2013qma} and hence fails to capture the long-term 
growth of the Einstein Rosen bridge of an AdS black hole. Thus, entanglement itself is not a suitable 
candidate to capture the dynamics behind the horizon \cite{Susskind:2014moa}. In order to capture the subsequent
dynamics, Susskind \emph{et al.} proposed that this late-time evolution of a wormhole is encoded in 
the quantum complexity of the boundary state. Several proposals for the holographic dual of the circuit 
complexity of the boundary state were put forward, such as the ``complexity=volume'' conjecture 
\cite{Susskind:2014rva,Stanford:2014jda} and the ``complexity=action" conjecture \cite{Brown:2015bva}. 

Due to its significance in holography, it is also crucial to quantify 
complexity in quantum field theory. In particular, the complexity of the ground states in scalar field theories \cite{Jefferson:2017sdb,Bhattacharyya:2018bbv}, thermofield double states \cite{Chapman:2018hou}, and squeezed vacuum states \cite{Lehners:2020pem, Bhattacharyya:2020rpy, Bhattacharyya:2020kgu, Bhargava:2020fhl} 
have been explored, and extended to fermionic cases in \cite{Khan:2018rzm, Hackl:2018ptj, Jiang:2018nzg}. The formalism developed to date is suitable for characterizations of the complexity of an operator that acts on target 
and reference states if they are Gaussian \footnote{Interested readers can refer to \cite{Chapman:2021jbh} for a review of the 
progress made and an extensive list of references.}. The complexity of Gaussian states has been studied using various 
approaches like Fubini-Study approach \cite{Chapman:2017rqy}, covariance matrix formalism \cite{Chapman:2018hou,Hackl:2018ptj,Ali:2018aon,Adhikari:2021ked}\footnote{A covariance-matrix formalism to compute the complexity of non-Gaussian states was extended in \cite{Guo:2020dsi}.}
or directly working with the wavefunctions in the position basis \cite{Jefferson:2017sdb,Bhattacharyya:2018bbv, Alves:2018qfv, Guo:2018kzl}. A comparative analysis between the 
different approaches to state complexity was done in \cite{Ali:2018fcz}. Such a restriction on the 
target and reference state is not suitable for making comments about general target unitaries, 
which are crucial in particular for interacting systems. The group manifold approach of quantum complexity 
allows one to generalize Nielsen's geometric method to general unitaries without making any connection to 
reference and target states \cite{Balasubramanian:2019wgd,Haque:2021hyw, Chowdhury:2023iwg}. 
This approach, therefore, provides a potential way to study quantum complexity in interacting systems and generalize the notion of 
complexity beyond Gaussian states. Furthermore, Nielsen's notion of complexity 
is particularly relevant to quantum computing in ways that other measures, such as Krylov complexity \cite{Caputa:2021sib}, are not. Nielsen introduced the geometric approach to complexity for 
$n$-qubit unitaries to determine the optimal circuit leading to a given unitary. This makes Nielsen's complexity directly applicable to quantum computing, as it focuses on minimizing the number of gates 
required to implement a desired unitary. Consequently, the geometric framework is not only theoretically significant but also motivated by practical experimental considerations.

In the group manifold approach, the Lie algebra generated by the fundamental operators that construct the target unitary operator is taken into consideration. The minimal geodesic in 
the corresponding Lie group manifold (upon a suitable choice of metric) is obtained by solving the Euler-Arnold 
equation. The advantage of this approach is that the geometry is entirely determined by the generators of the 
corresponding Lie algebra and, hence can be applied to interacting and anharmonic systems by suitably 
generalizing the Lie algebra. 

In the present paper, we apply the group manifold approach to study the complexity of the generator of time evolution for a
scalar field mode in de Sitter spacetime. This analysis will, in particular, allow us to quantify how difficult it
is to simulate the dynamics on future quantum computers. We begin with the general framework of the time evolution 
of a harmonic oscillator with time-dependent parameters. This quantum system is ubiquitous in most branches of physics and has been widely studied in numerous contexts. 
Dynamics of quantum fields in an expanding universe \cite{Birrell:1982ix,Mukhanov:2007zz}, dissipation in 
open quantum systems, and squeezing in quantum optics are a few such processes that can be efficiently 
understood using the model of a harmonic oscillator with time-dependent parameters. 

Some previous works have been done in this context \cite{Lehners:2020pem,Bhattacharyya:2020rpy, Bhattacharyya:2020kgu}. 
In \cite{Bhattacharyya:2020rpy,Bhattacharyya:2020kgu}, the complexity of scalar cosmological perturbations 
was explored on an expanding FRW background. Since the authors adopted the state-based complexity approach, 
a reference and a target state had to be selected. The ground state of the mode inside the horizon was 
chosen as the reference state (two-mode vacuum state), and the complexity of the target state consisted of 
the time-evolved cosmological perturbation on the expanding background (two-mode squeezed vacuum state). 
It was found that the complexity of cosmological perturbations in the de Sitter background is extremely 
small until the mode is inside the horizon (subhorizon modes), then grows linearly with the log of 
the scale factor and is thus proportional to the number of $e$-folds. Similarly, in \cite{Lehners:2020pem}, 
the quantum complexity of cosmological perturbations was investigated in different models of the 
early universe.  A comparative analysis was made between the models (including inflation, ekpyrosis, 
and a contracting matter phase). It was found that the matter-dominated contracting phase shows the most 
rapid growth in complexity while inflationary perturbations showed the least growth, with ekpyrotic
perturbations growth in between. 

The organization of this article is as follows: The primary approaches to determine the complexity of a 
unitary operation are reviewed in Sec.~\ref{sec2}. The main ideas behind the geometrical interpretation 
of the quantum complexity of a unitary operation and ways to extend Nielsen's idea to deal with systems 
having an infinite dimensional Hilbert space are discussed. The desired target unitary operator, which 
is the time evolution operator of an oscillator with time-dependent parameters, is established as a 
product of two unitary operators, popularly known as the squeezing and the rotation operator in Sec.~\ref{sec3}. In Sec.~\ref{sec4}, the complexity of the target unitary operator is determined, first 
geometrically following Nielsen's approach in Sec.~\ref{subsec:Nielsen} and then the approach of gate 
counting in Sec.~\ref{subsec:gatecomplexity}. The results obtained in the previous sections are illustrated 
with a simple example of an oscillator with a sudden frequency jump in Sec.~\ref{sec5}. The complexity 
of the time evolution of a scalar field mode in de Sitter spacetime is discussed in Sec.~\ref{sec6}. 
Section~\ref{sec7} summarizes the key results and conclusions of the paper.

\section{Insights into quantum complexity: A rapid review}
\label{sec2}

\subsection{The discrete picture: Gate complexity}
\label{subsec:gatecomplexity}

Complexity measures the minimum number of simple operations required to carry out a task. In the 
quantum circuit picture, we might think of complexity as the minimum number of elementary gates 
required to construct the quantum circuit that produces the desired target unitary that transforms 
a given reference state $\ket{\Psi}_R$ to a target state $\ket{\Psi}_T$ \emph{i.e}.: 
\begin{align}
    \ket{\Psi}_T= U\ket{\Psi}_R.
\end{align}
The next step involves identifying a simple set of unitary gates $\{g_i\}$ with which to construct 
$U$ as a product of $g_i's$:
\begin{align}
    U_{\rm target}= g_1g_2....g_i \mathbb{I}.
\end{align}

One can then define something known as the ``circuit depth'' as the total number of gates in 
the circuit. However, one must be careful enough to distinguish it from the complexity of $U$, which 
is the minimum number of gates required to construct $U$. In other words, complexity may be 
thought of as the circuit depth of the optimal circuit.

Let us understand it with the help of a simple example. Consider the case where we are interested 
in the complexity of a certain $U$. The first step requires identifying a simple set of unitary 
gates with which to construct $U$. Let us assume that for the target unitary operator, the gates 
$g_1$, $g_2$, and $g_3$ form the universal set:
\begin{align}
    g_1 = e^{i \epsilon J_1}, ~~~~ g_2 = e^{i \epsilon J_2}, ~~~~ g_3=e^{i \epsilon J_3}.
\end{align}

The infinitesimal parameter $\epsilon \ll 1$ ensures that the action of 
any gate $g_i$ produces only a small change on the identity operator 
$\mathbb{I}$. The $J_i$ are the generators that are exponentiated 
to form a gate. A general circuit that constructs $U$ then consists 
of a sequence of these gates, \emph{i.e.}: 
\begin{align}
\label{circuit}
    U= g_3^{n_3}g_2^{n_2}g_1^{n_1}.
\end{align}

The above equation indicates that the gate $g_1$ acts first an 
appropriate number of times, $n_1$. The \textit{circuit depth} 
can then be defined as the total number of gates in the circuit. 
In our example, we simply have: 
\begin{align}
\label{circuitdepth}
    \mathcal{D}[U_{\rm target}]= |n_1|+|n_2|+|n_3|.
\end{align}
The absolute values in the above line indicate that the inverse 
gate $g_i^{-1}$ is also considered, and the appearance of $g_i^{-1}$ 
is counted as one gate in a circuit. Equal complexity cost is given 
for the inverse gate $g_i^{-1}$ as for the original gate $g_i$.

Readers are referred to Refs. \cite{Jefferson:2017sdb,Lehners:2020pem} for illustrations with various 
reference and target wave functions. The result in Eq.~(\ref{circuitdepth}) is understood as the circuit 
depth of the specific circuit $U$. We will refer to this as \textit{gate complexity}. However, it is 
important to distinguish this from the actual complexity of $U_{\rm target}$, which is the minimum number 
of gates required to construct $U_{\rm target}$. In other words, the complexity is the circuit depth of 
the optimal circuit. We have no reason to believe that the circuit proposed in Eq.~(\ref{circuit}) is always 
the optimal one. In general, there may be a possibility of finding different gate sets that allow 
$U_{\rm target}$ to be constructed with fewer gates. However, determining an optimal circuit is challenging. 
Hence, the gate definition of complexity might not be very useful in determining the true difficulty 
of a computation. In the next subsection, we will discuss Nielsen's approach to geometrizing circuit 
complexity to find the optimal circuit.

\subsection{The continuous picture: Geometrizing quantum complexity}
\label{subsec:geometrycomplexity}

Nielsen \textit{et al.\ }\cite{Nielsen_2006,https://doi.org/10.48550/arxiv.quant-ph/0502070,https://doi.org/10.48550/arxiv.quant-ph/0701004} established 
new connections between differential geometry and quantum complexity in
several studies proposing a shift from the discrete explanation of gate
complexity to a continuous one. They observed that the problem of determining 
the quantum complexity of a unitary operation is related to the problem of 
finding minimal-length geodesics in certain curved geometries. A geometrical 
definition of quantum complexity was first proposed as a tool to constrain 
the value of gate complexity; from there, it developed into a contender for 
a distinctive definition of quantum complexity.

In the geometrical framework, the length of the minimal geodesic on the 
unitary group manifold connecting the identity to $U$ is the complexity 
of a unitary operator in this approach. The minimal geodesic corresponds
to the optimal circuit. An extension of the basic idea of the geometrical 
framework, which was based on unitaries acting on $n$-qubit systems to 
a general unitary is initially straightforward, but it does lead to several 
mathematical subtleties, some of which are described in Ref.~\cite{Chowdhury:2023iwg}. 

One begins by identifying the set of fundamental operators related to the
target unitary operator and classifying them as ``easy'' or ``hard''. 
To define a geometry, one then considers a right-invariant metric ($G_{IJ}$), 
popularly known as the \textit{penalty factor matrix} that accurately penalizes 
the directions along the hard operators such that moving in their 
direction is discouraged for geodesics in the Lie group. The choice of 
the matrix $G_{IJ}$ is usually motivated by phenomenological considerations, 
inspired by difficulties of performing certain operations \cite{Brown:2019whu}. 
The metric then leads to a notion of distance on the space of unitaries, 
which is given by
\begin{widetext}
\begin{align}
\label{lineelement}
    ds^2 = \frac{1}{{\rm Tr}(\mathcal{O}_I\mathcal{O}^{\dagger}_I) {\rm Tr}(\mathcal{O}_J\mathcal{O}^{\dagger}_J)}\bigg[G_{IJ} {\rm Tr}[i U^{-1}\mathcal{O}_I^{\dagger}dU] {\rm Tr}[i U^{-1}\mathcal{O}_J^{\dagger}dU]\bigg]\,.
\end{align}
\end{widetext}
The $\mathcal{O}_I$ represent the generators of the unitary group 
and $U$ plays the role of a point on the manifold. The trace ${\rm Tr}$ is 
taken in a matrix representation of the generators. For geodesics, only 
the right-invariance of the line element matters, but not the specific 
form on the entire group.

An efficient way of determining geodesics on Lie groups equipped with a 
right invariant metric was given by Arnold and is known as the Euler--Arnold equation \cite{Balasubramanian:2019wgd}. It has been extensively 
used recently to compute the geodesics on unitary manifolds \cite{Chowdhury:2023iwg,Haque:2021hyw,Balasubramanian:2019wgd,
Balasubramanian:2021mxo,Flory:2020dja}.  If  $f_{IJ}^{K}$ are the structure constants of the Lie algebra, defined by
\begin{equation}
\label{structure}
	[\mathcal{O}_{I},\mathcal{O}_{J}]= i f_{IJ}^{K} \mathcal{O}_{K}\,,
\end{equation}
the Euler--Arnold equation reads \footnote{Please refer to \cite{Haque:2024ldr} for a derivation of the Euler-Arnold equation as written in \ref{eqn:eulerarnoldrev}.}
\begin{equation}
\label{eqn:eulerarnoldrev}
	G_{IJ}\frac{dV^{J}(s)}{ds}= f_{IJ}^{K} V^{J}(s) G_{KL}V^{L}(s)\,.
\end{equation}

The components $V^I(s)$ represent the tangent vector (or the velocity) at each point along the geodesic, defined by:
\begin{align}
\label{differentialU}
	\frac{dU(s)}{ds}=-i V^I(s)\mathcal{O}_I U(s).
\end{align}
Given a solution $V^I(s)$, further integration of (\ref{differentialU}) 
results in the path (or trajectory) in the group, guided by the velocity 
vector $V^I(s)$. Generically, this solution can be written as the 
path-ordered exponential:
\begin{equation}
\label{sdependentunitary}
	 U(s)=\mathcal{P}\exp\bigg(-i\int_{0}^{s}ds'~V^I(s')\mathcal{O}_I\bigg),
\end{equation}
on which we impose the boundary conditions: 
\begin{equation}
\label{boundary}
	U(s=0)=\mathbb{I} ~~~~{\rm and}~~~ U(s=1)=U_{\rm target},
\end{equation}
where $U_{\rm target}$ is some target unitary whose complexity we wish to study. 

In general, equation (\ref{eqn:eulerarnoldrev}) defines a family of 
geodesics $\{V^I(s)\}$ on the unitary space. The boundary condition
$U(s=1)=U_{\rm target}$ selects those geodesics that can realize the 
target unitary operator by fixing the magnitude of the tangent vector 
$V^I$ at $s=0$ (at the identity operator). There could be more than 
one value of the $V^I$ for which the point of the target unitary is 
reached. If this is the case, the smallest value of the corresponding length along the curve $U(s)$ is to be considered as we are 
looking for the geodesic that minimizes the distance between two elements of the group. Therefore, the complexity 
can be defined as
\begin{equation}
\label{eqn:complexityexpression}
	C[U_{\rm target}] := {\rm min}_{\{V^I(s)\}}\int_{0}^{1}ds\sqrt{G_{IJ}V^{I}(s)V^{J}(s)},
\end{equation}
where the minimization is over all solutions $\{V^I(s)\}$ of the Euler--Arnold equation, providing extremal distances from 
the identity to the target unitary $U_{\rm target}$.

\section{Evolution operator of a time-dependent oscillator}
\label{sec3}

In this section, we will identify the desired target unitary operator 
whose quantum complexity we are interested in. For our purpose, 
since our main concern is the complexity of time evolution, our 
target unitary operator should be the time evolution operator of a given system. In 
the following, we will discuss the evolution operator of a 
time-dependent oscillator. 

Let us begin by considering a harmonic oscillator 
with time-dependent frequency $\omega(t)$, fixing the mass $m$ to 
equal 1 for the sake of convenience. (The latter choice can always be achieved by a canonical transformation of the basic variables $(q,p)$.) The Hamiltonian of such an oscillator can be written as
\begin{align}
\label{Hamiltonian}
    H(t)= \frac{p^2}{2}+\frac{1}{2}\omega^2(t)q^2,
\end{align}
where $p=\dot{q}$ and $q$ satisfies the second-order equation of motion
\begin{align}
\label{eomq}
    \Ddot{q}(t)+\omega^2(t)q(t)=0\,.
\end{align}
The canonically conjugated variables $q$ and $p$ can be promoted to operators:
\begin{align}
\label{qmode}
    {q}(t) &= f(t){a}_0+f^*(t){a}^{\dagger}_0, ~~~ {p}(t)= g(t) {a}_0+ g^*(t) {a}^{\dagger}_0\,,
\end{align}
where $a_0$ and $a^{\dagger}_0$ are the annihilation and creation operators defined at 
some initial time $t_0$. The mode functions $f(t), g(t) \in \mathbb{C}$ obey the condition:  
\begin{equation} \label{gdotf}
g(t)=\Dot{f}(t)
\end{equation}
as a direct consequence of $p=\dot{q}$. Since ${q}(t)$ satisfies Eq.~(\ref{eomq}), this is also the case 
for the mode function $f(t)$:
\begin{align} \label{ddotf}
\Ddot{f}(t)+\omega^2(t)f(t)=0.
\end{align}

The canonical commutation relation between the position 
and the momentum operator, $[{q},{p}]=i$, along
with the commutation relation satisfied by the ladder 
operators, $[{a},{a}^{\dagger}]=1$, leads to the 
Wronskian condition 
\begin{align}
    fg^{*}-f^*g=i\,.
\end{align}
Equations~(\ref{gdotf}) and (\ref{ddotf}), with real $\omega(t)$, imply that this condition holds at all times if it is imposed on initial values.
The time evolution of the creation and the annihilation operator gives us the system's time evolution. The annihilation and the creation operator at any time $t$ can be written as: 
\begin{align}
\label{bogoliubovtransformation}
    {a}(t)= \alpha^*(t){a}_0-\beta^*(t) {a}_0^{\dagger}\quad,\quad {a}^{\dagger}(t)= -\beta(t) {a}_0+\alpha(t){a}^{\dagger}_0, 
\end{align}
where $\alpha, \beta \in \mathbb{C}$, are the so-called Bogoliubov coefficients. 

Using two sets of creation and annihilation defined at two  
regimes, labeled as ``in'' and ``out'', the position operator ${q}$ can be written as:
\begin{align}
    {q}(t)= f(t){a}_{\rm in}+f^*(t){a}^{\dagger}_{ \rm in}= \Tilde{f}(t){a}_{\rm out}+\Tilde{f}^*(t){a}^{\dagger}_{\rm out}.
\end{align}
Now, using the Bogoliubov transformation, we can express ${a}_{\rm out}$ in terms of ${a}_{\rm in}$: 
\begin{widetext}
\begin{align}
    f(t){a}_{\rm in}+f^*(t){a}^{\dagger}_{\rm in}= \Tilde{f}(t)(\alpha^*(t){a}_{\rm in}-\beta^*(t) {a}_{\rm in}^{\dagger})+\Tilde{f}^*(t)(-\beta(t) {a}_{\rm in}+\alpha(t){a}^{\dagger}_{\rm in}).
\end{align}
\end{widetext}
Equating the coefficients of ${a}_{\rm in}$ and ${a}_{\rm in}^{\dagger}$, we obtain
\begin{align}
\label{feqns}
    f(t) &= \Tilde{f}(t)\alpha^*(t)-\Tilde{f}^*(t)\beta(t), \\
    f^*(t) &= -\Tilde{f}(t)\beta^*(t)+\Tilde{f}^*(t)\alpha(t).
\end{align}
Similarly, the momentum operator can be written as: 
\begin{align}
    {p}(t)= g(t){a}_{\rm in}+g^*(t){a}^{\dagger}_{\rm in}= \Tilde{g}(t){a}_{\rm out}+\Tilde{g}^*(t){a}^{\dagger}_{\rm out}.
\end{align}
Equating the coefficients of ${a}_{\rm in}$ and ${a}_{\rm in}^{\dagger}$, we obtain
\begin{align}
\label{geqns}
    g(t) &= \Tilde{g}(t)\alpha^*(t)-\Tilde{g}^*(t)\beta(t), \\
    g^*(t) &= -\Tilde{g}(t)\beta^*(t)+\Tilde{g}^*(t)\alpha(t).
\end{align}
Solving (\ref{feqns}) and (\ref{geqns}) and using the Wronskian condition: 
\begin{align}
    fg^{*}-f^*g= \Tilde{f}\Tilde{g}^*-\Tilde{f}^*\Tilde{g}=i,
\end{align}
we finally obtain expressions for the Bogoliubov coefficients 
being functions of the ``in'' and ``out'' mode functions: 
\begin{align}
\label{bogoliubov}
    \alpha & = \frac{\Tilde{f}g^*-f^* \Tilde{g}}{-\Tilde{f}^*\Tilde{g}+\Tilde{f}\Tilde{g}^*}= -i(\Tilde{f}g^*-f^* \Tilde{g}), \\
    \beta &= \frac{\Tilde{f}g-f\Tilde{g}}{\Tilde{f}^*\Tilde{g}-\Tilde{f}\Tilde{g}^*} = i (\Tilde{f}g-f\Tilde{g}).
\end{align}
Using the Wronskian condition, one can also verify that the above equations satisfy the normalization condition:
\begin{align}
    |\alpha|^2-|\beta|^2=1,
\end{align}
which guarantees that the commutation relation between the creation 
and the annihilation operator is preserved in time.

The above equation indicates that the Bogoliubov coefficients can be parametrized hyperbolically as: 
\begin{align}
\label{parametrizer}
    \alpha(t)= e^{-i \theta(t)}\cosh(r(t)), ~~ \beta(t) = e^{-i(\phi(t)-\theta(t))}\sinh(r(t)).
\end{align}

Using this parametrization, (\ref{bogoliubovtransformation}) can be written as: 
\begin{align}
\label{bogoliubivsqueezed}
    {a}(t) &= e^{i \theta(t)}\cosh(r(t)) {a}_0- e^{i(\phi(t)-\theta(t)}\sinh(r(t)) {a}^{\dagger}_0, \\
    {a}^{\dagger}(t) &= e^{-i \theta(t)}\cosh(r(t)) {a}_0- e^{-i(\phi(t)-\theta(t))}\sinh(r(t)) {a}^{\dagger}_0.
\end{align}

\subsection{Setting up the target unitary operator}
The Bogoliubov transformation can be represented as a similarity transformation,
\begin{align}
    {a}(t)= U^{\dagger}(t){a}(t_0) U(t)
\end{align}
for a unitary operator $U(t)$. 
In order to derive the form of $U$ corresponding to (\ref{bogoliubivsqueezed}), let us, for convenience, introduce two operators, which are known 
as the squeezing and the rotation operators, expressed as
\begin{align}
    {S}(\xi(t)) &= \exp\bigg(\frac{1}{2}(\xi^*(t) {a}^2-\xi(t) {a}^{\dagger 2})\bigg), \\
    {R}(\theta(t)) &= \exp\bigg(i \theta(t) \frac{a^{\dagger}a+a a^{\dagger}}{2}\bigg),
\end{align}
where $\xi(t)=r(t)e^{i\phi(t)}$.

The transformations of the creation and annihilation operators generated by $S$ and $R$ can be written as: 
\begin{align}
    S^{\dagger}(r(t),\phi(t))~{a}~S(r(t),\phi(t)) &= {a}\cosh(r(t)) \nonumber \\
    &-e^{i\phi(t)}{a}^{\dagger} \sinh(r(t)),\\
    R^{\dagger}(\theta(t)){a}R(\theta(t)) &= e^{i \theta(t)}{a}.
\end{align}
The derivation of the above equations has been relegated to Appendix \ref{appB}.
Considering (\ref{bogoliubivsqueezed}), it is evident that the transformation can be written as:
\begin{align}
\label{Canonicalform1}
    {a}(t) &= R^{\dagger}(\theta(t))S^{\dagger}(r(t),\phi(t)) ~{a}_0~ S(r(t),\phi(t))R(\theta(t)),
    \\
    \label{canonicalform2}
     {a}^{\dagger}(t) &= R^{\dagger}(\theta(t))S^{\dagger}(r(t),\phi(t)) ~{a}^{\dagger}_0~ S(r(t),\phi(t))R(\theta(t)).
\end{align} 

Therefore, the target unitary operator in this case is the product of the unitary operators $S$ and $R$:
\begin{align}
    U_{\rm target}= S(r(t),\phi(t))R(\theta(t)).
\end{align}
The time evolution of the parameters $r$ and $\phi$ can be determined from 
their relation with the Bogoliubov coefficients. In the following section, 
we will compute the complexity of the target unitary operator following 
a representation-independent approach discussed in Ref. \cite{Chowdhury:2023iwg}. 

\section{Complexity of the desired target unitary operator}
\label{sec4}

Our time evolution operators can, at any time, be written as the product of a squeezing and a rotation operator. It turns out the squeezing operator is more challenging in complex computations. (Heuristically, it is parameterized by a complex number $\xi$, which requires two non-commuting Hermitian generators in a Lie algebra. The real parameter $\theta$ of the rotation operator can instead be represented using a single Hermitian generator.) 

We, therefore begin with the squeezing operator,
\begin{align}
\label{squeezedoperator}
    S(\xi(t)) &= \exp\bigg(\frac{1}{2}(\xi^*(t) {a}^2-\xi(t) {a}^{\dagger 2})\bigg)
\end{align}
where $\xi(t)= r(t) e^{i \phi(t)}$, and rewrite it in a form suitable for geometrical methods.
The representation-independent procedure of \cite{Chowdhury:2023iwg} requires us to find a set of Hermitian operators 
that can be used to build (\ref{squeezedoperator}) and is closed with respect to taking commutators. It is not hard to check that the operators
\begin{align}
    \mathcal{O}_1= \frac{{a}^2+ {a}^{\dagger 2}}{4}, ~~~ \mathcal{O}_2= \frac{i({a}^2- {a}^{\dagger 2})}{4}, ~~~ \mathcal{O}_3 = \frac{{a}{a}^{\dagger}+ {a}^{\dagger}{a}}{4},
\end{align}
satisfy the commutation relations 
\begin{align}
\label{commutationrelations}
    [\mathcal{O}_1,\mathcal{O}_2]= -i \mathcal{O}_3, ~~~ [\mathcal{O}_1,\mathcal{O}_3]= -i \mathcal{O}_2,~~~ [\mathcal{O}_2,\mathcal{O}_3]= i \mathcal{O}_1\,,
\end{align}
forming the $\mathfrak{su}(1,1)$ Lie algebra. 
In terms of these generators, the squeezing operator can be 
written as: 
\begin{align}
\nonumber
    {S}(\xi) &= \exp\bigg(\frac{1}{2}(\xi^*(t) {a}^2-\xi(t) {a}^{\dagger 2})\bigg)\\ \nonumber
    &= \exp\bigg(\xi^*(t)(\mathcal{O}_1-i\mathcal{O}_2)-\xi(t)(\mathcal{O}_1+i\mathcal{O}_2)\bigg) \\
    &= \exp \bigg(-2ir(t)(\sin(\phi(t))\mathcal{O}_1+\cos(\phi(t))\mathcal{O}_2)\bigg).
\end{align}
The rotation operator can also be written in terms of the generators and takes the simple form
\begin{align}
    R(\theta(t))= \exp\bigg(i\theta\bigg(\frac{a^{\dagger}a+a a^{\dagger}}{2}\bigg)\bigg)= \exp(2i \theta(t) \mathcal{O}_3).
\end{align}

\subsection{Complexity via the geometric approach}
\label{subsec:Nielsen}

Therefore, the target unitary operator in terms of the generators 
can be written as: 
\begin{widetext}
\begin{align}
    S(r(t),\phi(t))R(\theta(t))= \exp \bigg(-2ir(t)(\sin(\phi(t))\mathcal{O}_1+\cos(\phi(t))\mathcal{O}_2)\bigg)\exp(2i \theta(t) \mathcal{O}_3).
\end{align}
\end{widetext}
As required, the generators $\mathcal{O}_I$ satisfy a closed commutator 
algebra and hence specify a Lie group by exponentiation. Geodesics in the 
corresponding Lie group manifold can be obtained from the 
Euler--Arnold equation, which can be written as 
\begin{equation}
\label{eqn:eulerarnold}
	G_{IJ}\frac{dV^{J}(s)}{ds}= f_{IJ}^{K} V^{J}(s) G_{KL}V^{L}(s)
\end{equation}
with structure functions $f_{IJ}^K$ read off from (\ref{commutationrelations}).

As discussed in Sec \ref{sec2}, $G_{IJ}$ classifies the generators according to whether they are ``easy'' 
or ``hard'' to construct. Typically, the ``easy'' operators 
are the ones with less than $k$-body interactions for some $k$. 
For instance, in Ref. \cite{Chowdhury:2023iwg}, the authors considered 
the case of two coupled oscillators, where the operators involving 
two oscillator terms were assigned higher penalties compared to 
oscillators involving one oscillator term. Similarly, in the example 
of the anharmonic oscillator considered in the same paper, 
the higher-order operators were considered ``hard'' and were assigned 
extremely large penalties in a limiting procedure. In
the present case, the operators $\mathcal{O}_1$, 
$\mathcal{O}_2$ and $\mathcal{O}_3$ do not satisfy any distinctive or higher-order criteria  for which 
they should be assigned different penalties. We therefore choose $G_{IJ}=\delta_{IJ}$, such that the Euler--Arnold equations can 
be written for individual components of the tangent vector as 
\begin{align}
\label{EUeqns}
    \frac{dV^1}{ds}= -2 V^2 V^3, ~~~ \frac{dV^2}{ds}= 2 V^1 V^3, ~~~ \frac{dV^3}{ds}=0.
\end{align}
The general solutions to Eq.~(\ref{EUeqns}) are  
\begin{align}
    V^1(s) &= v_1 \cos(2 v_3 s)-v_2 \sin(2 v_3 s), \\
    V^2(s) &= v_1 \sin(2 v_3 s)+ v_2 \cos(2 v_3 s), \\
    V^3(s) &= v_3,
\end{align}
with integration constants $v_I$, $i=1,2,3$, determined by the condition that the target unitary is reached in the group manifold at $s=1$.
Since the tangent vector has a constant norm along a geodesic, the length of the curve from $s=0$ to $s=1$ equals $\sqrt{||V^I||^2}$. Provided we choose the solutions $v_I$ that imply the shortest length, in case of multiple solutions to the target condition, 
the complexity of the target unitary operator is given by 
\begin{align} \label{Cgen}
C[U_{\rm target}]= \sqrt{v_1^2+v_2^2+v_3^2}.
\end{align}

In this expression, the target unitary is specified implicitly in terms of the $v_I$. Finalizing the derivation of the complexity requires an explicit relationship between a given target unitary and the coefficients $v_I$, which is usually the most challenging part of this method because, to this end, we
need to know the geodesic for fixed boundary conditions 
$U(s=0)= \mathbb{I}$ and $U(s)= U_{\rm target}$; knowing the tangent vector $V^I$ along the geodesic is not sufficient. 
In general, the unitary along the geodesic path, starting at the identity and proceeding from there with a specific 
tangent vector $V^I(s)$, is given by the path-ordered exponential
\begin{align}
    U(s)= \mathcal{P} \exp\bigg(-i\int_0^s V^I(s') \mathcal{O}_I ds'\bigg),
\end{align}
which is a solution to the equation: 
\begin{align}
\label{differentialU1}
    \frac{dU(s)}{ds}=-i V^I(s)\mathcal{O}_I U(s),
\end{align}
that takes into account the non-commuting nature of the right-hand side of (\ref{differentialU1}) evaluated at different $s$.

Our task is to solve the path-ordered exponential and see what $U(1)$ looks like as a function 
of $v_I$. Using the boundary condition $U(1)=U_{\rm target}$ then determines the $v_I$ for a specified target unitary operator. 
However,  dealing with path ordering in the formal solution for $U(s)$ is a notoriously difficult problem. It is usually approached by using an 
iterative method. Using the initial value $U(0)=\mathbb{I}$, we have 
\begin{equation} \label{Uint}
 U(s)=U(0)+\int_0^s \frac{dU(s')}{ds'}ds' =\mathbb{I}-i \int_0^s V^I(s')\mathcal{O}_IU(s') ds'\,.
\end{equation}
The integrand still depends on the unknown $U(s)$, but close to the initial value, we can approximate $U(s)\approx\mathbb{I}$ in the integral. The resulting expression $U(s)\approx \mathbb{I}- i \int_0^s V^I(s')\mathcal{O}_Ids'$. Inserting this approximation back into the integrand of (\ref{Uint}) gives us a higher-order approximation, and so on by iteration.
The result gives as a \textit{Dyson series}:
\begin{widetext}
\begin{align}
    U(s)= \mathbb{I}-i \int_{0}^{s}V^I(s')\mathcal{O}_I ds' + (-i)^2 \int_{0}^{s} V^I(s') \mathcal{O}_I ds'\int_{0}^{s'}V^J(s'')\mathcal{O}_J ds''+\cdots,
\end{align}
\end{widetext}
for the path-ordered exponential.

For constant $V^I(s)$, the Dyson series is reduced to a simple exponential because $V^I\mathcal{O}_I$ then commute with one another at different $s$. In the present case, this property is realized if $v_1=0=v_2$, which allows us to express the rotation operator as a target unitary. However, the squeezing operator cannot be obtained in this simple form and requires approximations. In particular, our leading-order results will be reliable provided $v_1$ and $v_2$ are much smaller than $v_3$. Our results in what follows can then be expressed as a Taylor expansion in $v_1/v_3$ and $v_2/v_3$ or in suitable group parameters that amount to these ratios. Since higher-order terms in the Dyson series are implied by non-vanishing commutators of the generators, which are proportional to $\hbar$ in standard units, the Dyson series can be interpreted as an expansion in $\hbar$ (or a loop expansion in the language of quantum field theory).

For now, we will keep only the leading-order term in the Dyson series. This approximation implies that we are not following an exact geodesic, such that the length of our approximate curve overestimates the geodesic distance. We, therefore, obtain an \textit{upper bound} on complexity rather than the precise value. (Upper bounds on complexity were also provided in \cite{Craps:2022ese,Craps:2023rur}, based on different methods.) Explicit results in concrete applications will then allow us to determine whether our approximation is self-consistent. In broad terms, the leading-order approximation is reliable whenever evolution stays close to a single one of the generators chosen to embed the target unitary in a Lie group. 

Substituting the $V^I(s)$ obtained from solving the Euler--Arnold equation in the Dyson series, we can write 
\begin{align}
\nonumber
    \int_0^s V^I(s')\mathcal{O}_I ds' =& \frac{\mathcal{O}_1 v_1 \sin (2 s v_3)}{2 v_3}-\frac{\mathcal{O}_1 v_2 \sin ^2(s v_3)}{v_3} \\ &+\frac{\mathcal{O}_2 v_1 \sin ^2(s v_3)}{v_3} +\frac{\mathcal{O}_2 v_2 \sin (2 s v_3)}{2 v_3} \nonumber \\
    &+\mathcal{O}_3 s v_3.
\end{align}
As the leading-order term in the path-ordered exponential, this result implies
\begin{align} \label{Usingle}
\nonumber
    U(s) &\approx \exp\bigg(-i\bigg(\bigg\{\frac{v_1 \sin(2 s v_3)}{2 v_3}-\frac{v_2 \sin^2(s v_3)}{v_3}\bigg\}\mathcal{O}_1 \\ \nonumber &~~~~~~~~~+\bigg\{\frac{v_2 \sin(2 s v_3)}{2 v_3}+\frac{v_1 \sin^2(s v_3)}{v_3}\bigg\}\mathcal{O}_2 \\ &~~~~~~~~~ +  s v_3 \mathcal{O}_3\bigg)\bigg).
\end{align}

The final boundary condition is given by:
\begin{widetext}
    \begin{align}
\label{originaltarget}
    U(s=1)=U_{\rm target}= \exp(-2i r(t)(\sin(\phi(t))\mathcal{O}_1+\cos(\phi(t))\mathcal{O}_2) \exp(2i\theta(t)\mathcal{O}_3),
\end{align}
\end{widetext}
which is given as the product of two exponentials rather than a single one, as in (\ref{Usingle}). 
Implementing the boundary condition, therefore, requires us to express the product of 
two exponentials appearing in the above equation as a single 
exponential. This can be done by using the Baker--Campbell--Hausdorff (BCH) formula: 
\begin{align}
    e^Xe^Y = e^Z,
\end{align}
where 
\begin{equation}
Z= X+Y+\frac{1}{2}[X,Y]+\frac{1}{12}[X,[X,Y]]-\frac{1}{12}[Y,[X,Y]]+\cdots
\end{equation}
Also, here, we will neglect the contributions coming from the nested 
commutators in a first approximation. Since we ignore contributions from terms related to commutators, such as $[X,Y]$, 
$[X,[X,Y]]$, and so on, the approximation can formally be viewed as an
expansion by powers of the structure constants.

As in the case of the Dyson series, this approximation amounts to an expansion
in $\hbar$, which appears linearly in all the structure constants if standard
units are used. In general, higher-order terms in $\hbar$ for the complexity
are a combination of $\hbar$-terms in the Dyson series and $\hbar$-terms from
the BCH formula. Approximating the Dyson series implies deviations of the
trajectory to the target unitary from the geodesic. The result is a
distance greater than the geodesic length (which by definition is always the
shortest distance) and, therefore, an upper bound on the
complexity. Approximating the BCH formula, by contrast, changes the target
unitary. Such an approximation might place the final operator closer to the
identity than desired and could, therefore, underestimate the distance in some
cases. The combination of the two leading-order contributions should, therefore,
be interpreted as an approximate upper bound. In Appendix~\ref{appC}, we
include an explicit example of a next-to-leading order term in the BCH formula
that does increase the upper bound inferred from the leading-order
term. Higher-order terms in the Dyson series will be considered elsewhere.

This approximation implies that the curves we consider do not reach the exact target unitary we are interested in. The difference between the desired
$U_{\rm target}$ and the actual $U_{\rm target}^{(1)}$ is obtained by comparing:
\begin{align}
    U_{\rm target}= e^Xe^Y, ~~~~~~~~~  U_{\rm target}^{(1)} \approx e^{X+Y},
\end{align}
and should be sufficiently small for a good approximation.
If we replace $U_{\rm target}^{(1)}$ with $U_{\rm target}$ in the present example, we obtain:
\begin{widetext}
\begin{align}
    U_{\rm target}^{(1)} \approx \exp\bigg(-2 i r(t) (\sin(\phi(t)\mathcal{O}_1+\cos(\phi(t)\mathcal{O}_2)+2 i \theta(t) \mathcal{O}_3\bigg). 
    \label{TargerUnitaryU}
\end{align}
\end{widetext}
The boundary condition $U(s=1)=U_{\rm target}^{(1)}$ then implies: 
\begin{align}
    v_3 &= -2 \theta(t),   \\
    \label{eqn1v}
    \frac{v_1 \sin(2 v_3)}{2 v_3}-\frac{v_2 \sin^2(v_3)}{v_3} & = 2 r(t) \sin(\phi(t)), \\
    \label{eqn2v}
    \frac{v_1 \sin^2(v_3)}{v_3}+\frac{v_2 \sin(2v_3)}{2v_3} & = 2 r(t) \cos(\phi(t)),
\end{align}
from the coefficients of $\mathcal{O}_3$, $\mathcal{O}_1$ and $\mathcal{O}_2$, respectively, in (\ref{TargerUnitaryU}).
These equations are solved by:
\begin{align}
   v_3 &= -2 \theta(t), \\ 
   \label{v1eqn}
   v_1 &= -4 \theta(t) r(t) \csc (2 \theta(t)) \sin (2 \theta(t) -\phi(t)), \\
   \label{v2eqn}
   v_2 &= 4 \theta(t) r(t) \csc (2 \theta(t)) \cos (2 \theta(t) -\phi(t)).
\end{align}

One must note that the equations (\ref{v1eqn}) and (\ref{v2eqn}), which 
are obtained from inverting equations (\ref{eqn1v}) and (\ref{eqn2v}), are not valid at 
$\theta= \pm \frac{\pi}{2}$. This limitation arose from the upper-bound approximation 
used in our analysis, where higher-order terms in the Dyson series were neglected when 
deriving the equations for $v_i$'s. The condition for reliable results from the leading order of the Dyson expansion, given by $v_1/v_3\ll1$ and $v_2/v_3\ll1$, is realized if $r(t)\ll1$ at all times, provided $\theta(t)$ is not small. If $\theta\ll1$, the condition on $r(t)$ has to be strengthened to $r(t)\ll\theta(t)$ in order to overcome large factors of $\csc(2\theta(t))$ in $v_1$ and $v_2$. If $\theta(t)=0$, $r(t)$ would also be restricted to vanish by the last inequality. However, in this case, constant $\phi(t)$ allows us to use our leading-order approximation for any $r(t)$ because $U_{\rm target}$ is then a time-dependent exponentiation of a fixed linear combination of the generators, given by $\sin(\phi)\mathcal{O}_1+\cos(\phi)\mathcal{O}_2$. The same conditions ensure that $U_{\rm target}^{(1)}$ is close to $U_{\rm target}$ because the latter is mainly given by the direct exponentiation of a single generator, $\exp(2i\theta(t)\mathcal{O}_3)$.

Using (\ref{Cgen}), the upper bound on the complexity  of $U_{\rm target}^{(1)}$ is therefore given by
\begin{align}
\label{upperbound}
C[U_{\rm target}^{(1)}] \lesssim 2  \sqrt{\theta(t)^2(1+4~r(t)^2 \csc ^2(2 \theta(t) ))}.
\end{align}
For $\theta(t)=0$, we obtain the special case of
\begin{align} \label{Cr}
C[U_{\rm target}^{(1)}] \lesssim 2 r(t)\,.
\end{align}
The upper bound on complexity does not depend on $\phi(t)$, but for $\theta(t)=0$ we have to restrict reliable cases to constant $\phi(t)$ because this condition was used in intermediate steps.
When the parameter $\theta(t)$ is small but non-zero, the expression 
simplifies to:
\begin{align}
\label{asymptoticsmall}
C[U_{\rm target}^{(1)}] \lesssim 2 \sqrt{\theta(t)^2+r(t)^2}\,.
\end{align}
Our reliability condition then requires
$r\ll\theta\ll1$, such that
\begin{equation}\label{Ctheta}
    C[U_{\rm target}^{(1)}] \lesssim 2 |\theta(t)| +\frac{r(t)^2}{|\theta(t)|}\,.
\end{equation}

More generally, the upper bound can be written in terms of the Bogoliubov 
coefficients by inverting the equations (\ref{parametrizer}) and realizing 
that the parameters $r$, $\theta$, and $\phi$ can be parameterized by
\begin{align}
    r= {\rm arsinh}|\beta|, ~~~ 
    \theta= -{\rm arg} (\alpha), ~~~~ \phi= -{\rm arg} (\alpha \beta).
\end{align} 
This allows us to rewrite the upper bound (\ref{upperbound}) as:
\begin{widetext}
\begin{align}
\label{complexityupperboundM}
C[U_{\rm target}^{(1)}] \lesssim 2\sqrt{{\rm arg}(\alpha(t))^2(1+4 
{\rm arsinh}^2|\beta(t)| 
   \csc^2(2 {\rm arg}(\alpha(t))))}.
\end{align}
\end{widetext}
In our examples, a useful approximation will be given by $\alpha$ close to one and real, as well as $\beta$ small and real. In this case, we have vanishing $\theta$ and $\phi$, such that we can use the approximation corresponding to (\ref{Cr}) and further expand in $\beta$:
\begin{align}
\label{complexityupperboundMsmallbeta}
C[U_{\rm target}^{(1)}] \lesssim 2
{\rm arsinh}|\beta(t)| \approx 2\beta.
\end{align}

\subsection{A gate complexity approach}
\label{subsec:gatecomplexity}

In this section, we will try to determine the complexity of the 
target unitary operator (\ref{originaltarget}) by adopting the 
approach of counting the number of gates. The primary step in this case, as 
discussed in Section~\ref{subsec:gatecomplexity}, 
is to identify a set of unitary gates from which the circuit is to be constructed. For the target unitary operator (\ref{originaltarget}), 
it is not difficult to guess that the generators $\mathcal{O}_1$, 
$\mathcal{O}_2$, and $\mathcal{O}_3$ can be used to build gates 
that will be sufficient for our purpose. Let us name those gates 
$g_1$, $g_2$ and $g_3$ respectively: 
\begin{align}
    g_1= e^{-i \epsilon \mathcal{O}_1}, ~~~ g_2= e^{-i \epsilon \mathcal{O}_2}, ~~~ g_3= e^{-i\epsilon \mathcal{O}_3}.
\end{align}
The infinitesimal parameter $\epsilon$ ensures that the action 
of these gates produces a small change in the identity operator. 
Furthermore, ignoring any $O(\epsilon^2)$ contributions to the exponents that may be generated by products of the gates, according to the BCH formula, implies 
that the gates $g_1$, $g_2$ and $g_3$  commute to this order. 

Under this assumption, a general circuit built from the gates 
$g_1$, $g_2$, and $g_3$ can be expressed as:  
\begin{align}
    U \approx g_1^{n_1}g_2^{n_2}g_3^{n_3},
\end{align}
with, $n_i \in \mathbb{Z}$, or 
\begin{align}
    U \approx e^{-i\epsilon n_1  \mathcal{O}_1}e^{-i \epsilon n_2  \mathcal{O}_2}e^{-i \epsilon n_3  \mathcal{O}_3}.
\end{align}
By relating this expression to the target unitary operator (\ref{TargerUnitaryU}), 
we obtain
\begin{align} \label{gateequality}
    e^{-i\epsilon(n_1 \mathcal{O}_1+n_2 \mathcal{O}_2+n_3 \mathcal{O}_3)} \approx e^{-2 i r\sin(\phi)\mathcal{O}_1-2i r\cos(\phi)\mathcal{O}_2+2i \theta \mathcal{O}_3},
\end{align}
which gives us the following conditions:
\begin{align}
    n_1&= \frac{2}{\epsilon} r(t) \sin(\phi(t)),\\
    n_2 &= \frac{2}{\epsilon} r(t) \cos(\phi(t)),\\
    n_3 &= -\frac{2}{\epsilon} \theta(t).
\end{align}

Consequently, the \textit{circuit depth} in this case can be written as: 
\begin{widetext}
\begin{align}
\label{circuitdepthum}
D[U_{\rm target}^{(1)}] &:= |n_1|+|n_2|+|n_3| = \frac{1}{\epsilon}\bigg[|2 r(t) \sin(\phi(t))|+|2 r(t) \cos(\phi(t))|+|2 \theta(t)| \bigg] \nonumber \\
&= \frac{1}{\epsilon}\bigg[\bigg|2 {\rm arsinh}|\beta|(\sin({\rm arg}(\alpha \beta))\bigg|+\bigg|2 {\rm arsinh}|\beta|\cos({\rm arg}(\alpha \beta)))\bigg|+|2{\rm arg}(\alpha)|\bigg].
\end{align}
\end{widetext}
For small values of $r, \theta$ and $\phi$, the above expression can be simplified as: 
\begin{align}
    \epsilon D[U_{\rm target}^{(1)}] \approx |2r(t)(\phi(t)+1)| +|2 \theta(t)|.
\end{align}

As discussed earlier, the number of gates in the circuit gives us the
\textit{circuit depth} and not necessarily the \textit{complexity}. For 
example, if instead of the considered gate set in the above derivation, 
we use a different one, the gate complexity might change.

Let us consider the gate set $\{g_2,g_3,g_4,g_5\}$, where the $g_i$'s are given by:
\begin{align}
    g_2 &= e^{-i\epsilon \mathcal{O}_2},~~  g_3= e^{-i\epsilon \mathcal{O}_3}, \\~~ g_4 &= e^{-i\epsilon \mathcal{O}_2\mathcal{O}_3},~~~ g_5= e^{-i\epsilon \mathcal{O}_3\mathcal{O}_2}.
\end{align}
The circuit can, therefore, be written as: 
\begin{align}
    U &= e^{-i\epsilon n_2 \mathcal{O}_2}e^{-i\epsilon n_3\mathcal{O}_3}e^{-i\epsilon n_4\mathcal{O}_2\mathcal{O}_3}e^{-i\epsilon n_5 \mathcal{O}_3\mathcal{O}_2}  \nonumber \\
    &\approx e^{-i\epsilon(n_2 \mathcal{O}_2+n_3 \mathcal{O}_3+n_4\mathcal{O}_2\mathcal{O}_3+n_5\mathcal{O}_3\mathcal{O}_2)} \nonumber \\
    & \approx e^{-i\epsilon(n_2 \mathcal{O}_2+n_3 \mathcal{O}_3+n_4\mathcal{O}_2\mathcal{O}_3+n_5(\mathcal{O}_2\mathcal{O}_3-i \mathcal{O}_1))} \nonumber \\
    & \approx e^{-i\epsilon n_2\mathcal{O}_2-i\epsilon n_3 \mathcal{O}_3-i\epsilon(n_4+n_5)\mathcal{O}_2\mathcal{O}_3-\epsilon n_5\mathcal{O}_1}.
\end{align}
This should be equal to $\approx e^{-2 i r \sin(\phi)\mathcal{O}_1-2i r\cos(\phi)\mathcal{O}_2+2i \theta \mathcal{O}_3}$, which gives: 
\begin{align}
    n_5 &=\frac{1}{\epsilon} 2i r \sin(\phi), ~~ n_2= \frac{1}{\epsilon}2 r \cos(\phi), \\ n_3 &= -\frac{1}{\epsilon}2\theta, ~~~ n_4+n_5=0.
\end{align}
which gives $n_4= -\frac{1}{\epsilon}2ir \sin(\phi)$. Accordingly, the 
gate complexity of this circuit is given by: 
\begin{widetext}
\begin{align}
    D= |n_2|+|n_3|+|n_4|+|n_5|= \frac{1}{\epsilon}|4 r \sin(\phi)|+\frac{1}{\epsilon}|2 r \cos(\phi)|+ \frac{1}{\epsilon}|2\theta|.
\end{align}
\end{widetext}
The formula differs from Eq. (\ref{circuitdepthum}) by the factor 
2 in front of the sinus term. This shows that the circuit depth 
of this circuit is more than the previous circuit we consider, 
suggesting that the previous circuit we considered was better, 
requiring a lesser number of gates to construct $U_{\rm target}$.

As seen in the prior analysis, the gate complexity result 
depends on the gate choice. A natural question thus arises as to which 
set of gates one should choose to construct the circuit. As discussed earlier, 
the gate complexity counts the minimal number of gates required to 
construct the desired unitary. It thus quantifies the amount of resources required 
to construct the unitary. Of course, a better choice would be the one that 
requires a lesser number and involves a lesser number to choose from. For instance, 
the gate set $\{g_1,g_2,g_3\}$ has only three gates to choose from whereas the set 
$\{g_2,g_3,g_4,g_5\}$ has four. Another way to justify the superiority of a particular 
gate set over another is its simplicity. The gate set $\{g_2,g_3,g_4,g_5\}$ has 
gates that involve the product of the generators $O_I$ unlike the other gate set 
$\{g_1,g_2,g_3,g_4\}$, which does not involve such terms. This argument is directly 
relevant to the choice of the ``penalty factor matrix'' in the geometrical 
approach, where the operators involving the product of generators $O_I$ are 
considered hard, and the corresponding direction in the operator space is 
penalized.

A noteworthy observation from this analysis is that, unlike 
geometric complexity, gate complexity varies with the squeezing angle $\phi$. 
Moreover, gate complexity is expected to serve as an upper bound to geometric 
complexity. Consequently, the gate complexity calculations presented in this 
section provide a solid foundation for performing a comparative analysis with 
the results obtained for geometric complexity.

\section{Application 1: An oscillator with switched frequency}
\label{sec5}

\begin{figure}[h!]
    \centering
    \includegraphics[scale=0.4]{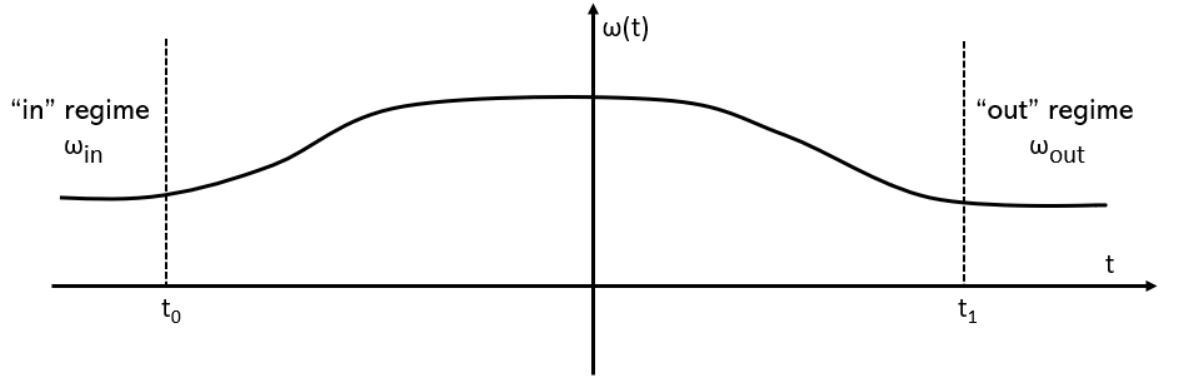}
    \caption{A sample frequency function $\omega(t)$ of the oscillator with constant frequency $\omega_{in}$ and $\omega_{out}$ in the ``in'' and ``out'' regimes.}
    \label{oscillatorfrequencyprofile}
\end{figure}

A popular model of a time-dependent oscillator is the case of a frequency that is constant 
in two disjoint time intervals, $\omega= \omega_{\rm in}$ for $t \leq t_0$ and 
$\omega=\omega_{\rm out}$ for $t\geq t_1$, but changes between $t_0$ and $t_1$. 
An example is shown in  Fig.~\ref{oscillatorfrequencyprofile}

We will be interested in relating the behavior of the oscillator in these two 
regimes, which are commonly called the \textit{in} and \textit{out} regimes. 
This model is noteworthy in the context of quantum quenching, 
where the thermalization properties of a system are analyzed following abrupt changes 
in the Hamiltonian. Furthermore, diverse applications of the model include 
cosmological particle production, owing to well-defined vacuum states in the 
two regimes.

In the intermediate region where $\omega(t)$ is not constant, an instantaneous 
ground state defined at time $t$ (defined by using the standard harmonic oscillator 
with a constant $\omega=\omega(t)$ at this fixed $t$, will not be a ground state at 
the next moment $t+\delta t$. Hence, such a ground state is not a physical one.
However, the ground state $\ket{0}_{\rm in}$ defined by the mode function $f_{\rm in}$
in the \textit{in} regime is a well-defined ground state for all $t\leq t_0$.
Similarly, the mode function $f_{\rm out}$ defines the ground state $\ket{0}_{\rm out}$
which is well-defined. The mode functions in the two regimes are given by: 
\begin{align}
\label{modein}
    f_{\rm in} &= \frac{1}{\sqrt{2\omega_{\rm in}}}e^{-i \omega_{\rm in}t} ~~~~ &{\rm for} ~ t\leq t_0, \\
    \label{modeout}
    f_{\rm out} &= \frac{1}{\sqrt{2\omega_{\rm out}}}e^{-i \omega_{\rm out}t} ~~~~ &{\rm for}~ t\geq t_1.
\end{align}
However, the mode function between $t_0$ and $t_1$ is not of this form and, in general, is not available in closed form. Nevertheless, the relationship between solutions in the \textit{in} and \textit{out} regimes can be analyzed by suitable approximations. 

\begin{figure}[h!]
    \centering
    \includegraphics[scale=0.5]{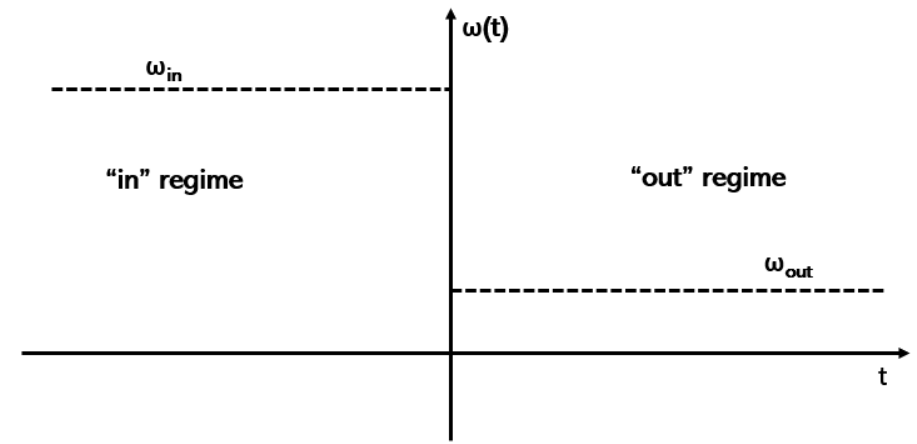}
    \caption{Diagrammatic representation of the frequency profile considered for the transition (\ref{inout}).}
    \label{suddenfrequencyprofile}
\end{figure}

To be specific, we consider a simple version of the above scenario in which the frequency experiences a sudden jump at $t=0$ from a constant frequency $\omega_{\rm in}$ to a different constant $\omega_{\rm out}$, as illustrated in Fig.~\ref{suddenfrequencyprofile}. The mode functions in the two regimes are given by equations (\ref{modein}) and (\ref{modeout}). Let $\ket{0}_{\rm in}$ and $\ket{0}_{\rm out}$ refer to the vacuum state of the oscillators before and after the frequency change, respectively. We are interested in the quantum complexity of $\ket{0}_{\rm out}$ with $\ket{0}_{\rm in}$ as the reference state, corresponding to the transition 
\begin{align} \label{inout}
    \underbrace{\ket{0}_{\rm in}}_{\text{reference state}} \longrightarrow \underbrace{\ket{0}_{\rm out}}_{\text{target state}}.
\end{align}
It is well known that this transformation can be written in terms of the squeezing operator,
\begin{align}
    \ket{0}_{\rm out}= {S}(\xi)\ket{0}_{\rm in}.
\end{align}
We can say that the new vacuum state is obtained by squeezing the old vacuum. Hence, the target unitary operator that we should be concerned about is the squeezing operator with the parameter $\xi= r e^{i \phi}$. The parameters $r$, $\theta$, and $\phi$ are related to the Bogoliubov coefficients as written in (\ref{parametrizer}).  

The ``out" mode functions can be written in terms of the ``in" mode functions using the Bogoliubov coefficients: 
\begin{align}
    f_{\rm out}(t) &= \alpha f_{\rm in}(t)+\beta f_{\rm in}^*(t),\\
    g_{\rm  out}(t) &= \alpha g_{\rm in}(t)+\beta g_{\rm in}^*(t).
\end{align}
Matching the conditions at $t=0$, we obtain 
\begin{align}
    \frac{1}{\sqrt{2\omega_{\rm out}}} &= \frac{ \alpha}{\sqrt{2\omega_{\rm in}}}+ \frac{\beta }{\sqrt{2\omega_{\rm in}}}, \\
    \sqrt{\frac{\omega_{\rm out}}{2}}  &= \sqrt{\frac{\omega_{\rm in}}{2}}\alpha -\sqrt{\frac{\omega_{\rm in}}{2}}\beta.
\end{align}
The Bogoliubov coefficients can then be calculated as 
\begin{align}
    \alpha= \frac{1}{2}\bigg(\sqrt{\frac{\omega_{\rm in}}{\omega_{\rm out}}}+\sqrt{\frac{\omega_{\rm out}}{\omega_{\rm in}}}\bigg), ~~~
    \beta = \frac{1}{2}\bigg(\sqrt{\frac{\omega_{\rm in}}{\omega_{\rm out}}}-\sqrt{\frac{\omega_{\rm out}}{\omega_{\rm in}}}\bigg).
\end{align}

\begin{figure}[h!]
    \centering
    \includegraphics[scale=0.5]{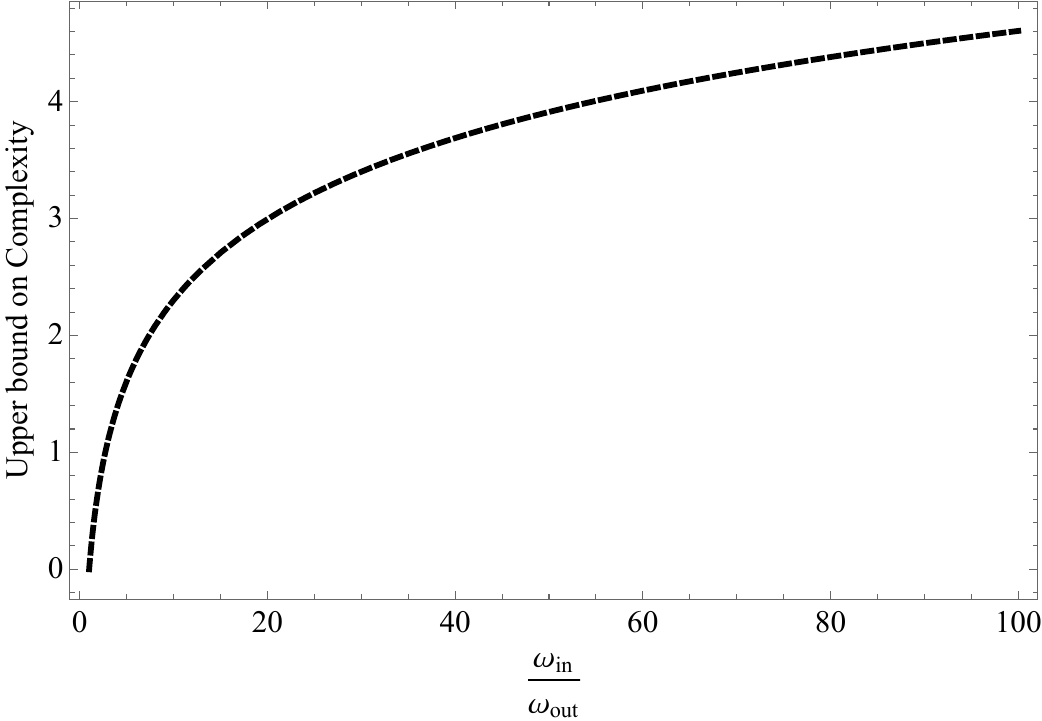}
    \caption{Behavior of the upper bound on complexity for an oscillator with switched frequency as a function of $\omega_{\rm in}/\omega_{\rm out}$.}
    \label{fig:switchedfrequencyoscillator}
\end{figure}

The upper bound on the complexity of the transformation from $\ket{0}_{\rm in}$  to $\ket{0}_{\rm out}$ 
is obtained by substituting the expression of the 
Bogoliubov coefficients in (\ref{complexityupperboundM}), which 
gives us
\begin{align}
\label{complexityoscswitch}
   C \lesssim 2{\rm arsinh}\bigg|\frac{1}{2}\bigg(\sqrt{\frac{\omega_{\rm in}}{\omega_{\rm out}}}-\sqrt{\frac{\omega_{\rm out}}{\omega_{\rm in}}}\bigg)\bigg|.
\end{align}
Since $\alpha$ and $\beta$ are both real, this case corresponds to vanishing $\theta$ and $\phi$ with a reliable leading-order term in the Dyson series.
Figure~\ref{fig:switchedfrequencyoscillator} shows how the complexity of an oscillator with switched frequency changes as a function of $\omega_{\rm in}/\omega_{\rm out}$. When $\omega_{\rm in}=\omega_{\rm out}$, $\ket{0}_{\rm in}=\ket{0}_{\rm out}$, such that the reference and target states coincide and hence the target unitary operator is the identity operator the complexity vanishes as expected. 

For an observer in the asymptotic future, the state $\ket{0}_{\rm in}$ is not a vacuum 
state but a state with particles. So if one builds a quantum circuit that simulates the $\ket{0}_{\rm out}$ 
state starting from the vacuum state $\ket{0}_{\rm in}$, the complexity of that circuit is upper bounded 
by (\ref{complexityoscswitch}). 
Equation (\ref{complexityoscswitch}) can also be expressed as: 
\begin{align}
    C \lesssim 2{\rm arsinh} \sqrt{n}, 
\end{align}
where $n=|\beta|^2$ is the number of particles. For small $n$, it can be 
approximated as:
\begin{align}
    C \lesssim 2 \sqrt{n}.
\end{align}
Interestingly, it is worth noticing that one can constrain the complexity of 
the produced state by measuring the number of particles produced.

\section{Application 2: Scalar field on de Sitter background}
\label{sec6}

In this section, we will deal with frequency profiles relevant to the context of quantum fields in curved spacetime. Each 
mode of a free quantum field on a non-static background behaves like a harmonic oscillator 
with time-dependent frequency. 

The action of a minimally coupled massive scalar field in curved spacetime 
is given by \cite{Mukhanov:2007zz,Birrell:1982ix}:
\begin{align}
\label{action}
    S= -\int d^4x \sqrt{-g}\bigg(\frac{1}{2}g^{\mu \nu}\partial_{\mu}\phi \partial_{\nu}\phi+ \frac{1}{2}m^2\phi^2\bigg).
\end{align}
Here, we will consider the special case of the flat 
Friedmann-Lemaitre-Robertson-Walker (FLRW) metric,
\begin{align}
    ds^2 = -dt^2+ a^2(t) d\vec{x}^2.
\end{align}
Introducing the notion of conformal time, $\tau$, related to the time 
coordinate $t$ by
\begin{align}
    \tau(t)= \int \frac{dt'}{a(t')},
\end{align}
the FLRW metric can be written as: 
\begin{align}
\label{metricFRW}
    ds^2= a^2(\tau)(-d\tau^2+d\vec{x}^2).
\end{align}
By introducing an auxilliary field $v= a(\tau)\phi$ and using (\ref{metricFRW}), 
the action (\ref{action}) takes the form 
\begin{align}
    S= \frac{1}{2}\int d\tau d^3x \bigg(\frac{1}{2}v'^2-\frac{1}{2}(\partial_i v)^2-\bigg(\underbrace{a^2m^2-\frac{a''}{a}}_{m_{\rm eff}^2}\bigg)v^2\bigg)
\end{align}
with the time-dependent effective mass $m_{\rm eff}$.
The field $v$ satisfies the equation
\begin{align}
\label{fieldequation}
    v''-\Delta v+ m_{\rm eff}^2 v=0.
\end{align}

Expanding the field $v$ in Fourier modes,
\begin{align}
    v(\vec{x},\tau)= \int \frac{d^3 \vec{k}}{(2 \pi)^{3/2}}v_{\vec{k}}(\tau)e^{i \vec{k}.\vec{x}}\,,
\end{align}
and substituting in the field equation (\ref{fieldequation}), 
we find the mode equation
\begin{align}
\label{modeEOM}
     v_{\vec{k}}''+ \bigg(\underbrace{k^2+a^2m^2-\frac{a''}{a}}_{\omega_k^2(\tau)}\bigg)v_{\vec{k}}=0.
\end{align}
Thus, a particular mode of a massive scalar field in FRW spacetime behaves like a harmonic oscillator 
with a time-dependent frequency given by
\begin{align}
    \omega_k^2(\tau)= k^2+a^2m^2-\frac{a''}{a}.
\end{align}
The time dependence is captured by the contribution from $m_{\rm eff}^2$, which encapsulates all the information about the influence of the gravitational background on the scalar field $\phi$.

For a quantum field, the general solution $v_{\vec{k}}(\tau)$ to the mode equation is expressed as a linear combination of creation and annihilation operators for the given wave number $\vec{k}$:
\begin{align}
    v_{\Vec{k}}(\tau)= f_k(\tau)a_{\Vec{k}}+ f_k^*(\tau) a_{-\Vec{k}}^{\dagger},
\end{align}
where the mode function $f_k(\tau)$ satisfies the classical equation,
\begin{align}
\label{modefunctioneqn}
    f_k''+\omega_k^2(\tau) f_k=0 , ~~~~~~~  \omega_k^2(\tau)=k^2+m_{\rm eff}^2.
\end{align}
We can, therefore, use the formalism developed in Section \ref{sec3}.

 A specific model that plays a significant role in cosmology is the 
case of de Sitter spacetime, which corresponds to the vacuum universe with 
positive cosmological constant $\Lambda$. Neglecting the curvature term 
the Friedmann equation takes the form $H^2=\frac{\Lambda}{3}$, which 
leads to the exponential solutions $a=a_0 e^{\pm H t}$, with a constant $a_0$.
The solution for the increasing branch parametrized by the conformal time 
$\tau$ takes the form  $a(\tau)=-1/(H \tau)$, where $\tau \in (-\infty, 0]$.

Consequently, the frequency profile of a single mode of a scalar field on 
expanding de Sitter background is given by: 
\begin{align}
\label{omegadeSitter}
    \omega_k^2(\tau)= k^2+\bigg(\frac{m^2}{H^2}-2\bigg)\frac{1}{\tau^2}.
\end{align}
The general solution of the mode function for this frequency profile can be expressed as:
\begin{align}
    f_k(\tau)= \sqrt{k|\tau|}(A J_n(k|\tau|)+B Y_n(k|\tau|)), ~~~ n=\sqrt{\frac{9}{4}-\frac{m^2}{H^2}}
\end{align}
with Bessel functions $J_n$ and $Y_n$.

Large values of $k|\tau|$ correspond to wavelengths that 
are much shorter than the Hubble distance $H^{-1}$ at time $\tau$. 
These are the \textit{subhorizon} modes, which are essentially 
unaffected by the curvature of spacetime. 
Conversely, small values of $k|\tau|$ correspond to physical 
wavelengths stretching far beyond the Hubble radius and are greatly 
affected by gravity. A mode with wavenumber $k$ is subhorizon 
at early times and becomes \textit{superhorizon} at a time 
($\tau=\tau_k)$ at which the physical wavelength is equal to 
the Hubble scale, $k|\tau_k|=1$. The time 
$\tau_k$ is referred to as the moment of \textit{horizon crossing}. 

For the time being, we will be interested in the massless
case, in which the frequency function is given by
\begin{align}
\label{frequencydesitter}
\omega_{dS}^2(\tau)= k^2- \frac{2}{\tau^2}.
\end{align}
The mode function, therefore, satisfies the equation
\begin{align}
    f_{k}''(\tau)+ \bigg(k^2-\frac{2}{\tau^2}\bigg)f_{k}=0
\end{align}
with general solution
\begin{align}
    f_k= -A_k \bigg(1-\frac{i}{k\tau}\bigg)\frac{e^{-ik\tau}}{\sqrt{2k}}-B_k \bigg(1+\frac{i}{k\tau}\bigg)\frac{e^{ik\tau}}{\sqrt{2k}}.
\end{align}
The constants $A_k$ and $B_k$, which determine the mode functions, should
chosen so as to obtain a suitable physically motivated vacuum state. 

For quantum fields in de Sitter spacetime, the preferred vacuum 
state is known as the \textit{Bunch-Davies} vacuum. Implementing this state
results in the mode functions
\begin{align}
\label{modedesitter}
  f_k (\tau) &= \frac{e^{-ik \tau}}{\sqrt{2k}}\bigg(1-\frac{i}{k\tau}\bigg), \\
  g_k(\tau) &=f_k' (\tau) = -i\sqrt{\frac{k}{2}}e^{-ik \tau} \bigg(1-\frac{i}{k\tau} -\frac{1}{(k\tau)^2}\bigg).
\end{align}
To obtain the Bogoliubov coefficients, we use Eq.~(\ref{bogoliubov}) 
in which $f$ is the Minkowski mode function and $\Tilde{f}$ 
the de Sitter one, 
\begin{align}
    f(\tau)=\frac{e^{-ik\tau}}{\sqrt{2k}}, ~~~~ \Tilde{f}(\tau)= \frac{e^{-ik \tau}}{\sqrt{2k}}\bigg(1-\frac{i}{k\tau}\bigg).
\end{align}

\begin{figure}
    \centering
    \includegraphics[scale=0.45]{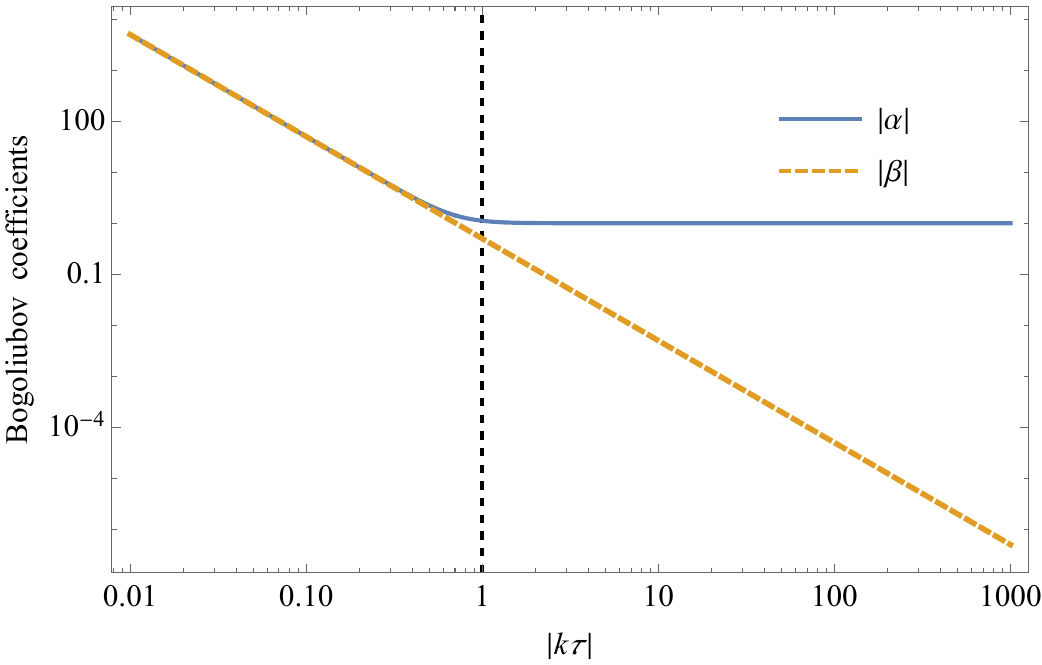}
    \caption{Log-log plot of the absolute values of the Bogoliubov
      coefficients (\ref{bogoliubovn}) as functions of $|k\tau|$. Conformal
      time $\tau$ takes negative values, but since it appears in an absolute value,
evolution for an expanding universe happens
from right to left in this plot. Large values of $|k\tau|$, therefore, correspond 
to initial time scales and vice versa.  At early times, $\tau \rightarrow -\infty$ or 
$|k\tau|\rightarrow \infty$, $|\beta| \rightarrow 0$ and $|\alpha| \rightarrow 1$. 
At late times, when $|k\tau| \rightarrow 0$, the absolute 
values of the Bogoliubov coefficients $\alpha$ and $\beta$ converge. 
The vertical dashed line separates the super-Hubble (left) and sub-Hubble (right) regions.}
    \label{fig:bogoliubovcoefficientplot}
\end{figure}

The transformation is understood such that in the limit 
$\tau\rightarrow -\infty$ any mode $k$ will be described by 
the Minkowski vacuum.  Therefore, the Minkowski vacuum can be 
considered as the ``in'' state.  At any further time $\tau$, we 
have the Bunch-Davies vacuum and the corresponding 
transformation between the Minkowski modes and the Bunch-Davies 
modes. This will lead to time-dependent expressions on the 
Bogoliubov coefficients, which are: 
\begin{align}
\label{bogoliubovn}
    \alpha(\tau) = 1- \frac{1}{2 k^2 \tau^2}-\frac{i}{k \tau}, ~~~~~
    \beta(\tau) = \frac{e^{-2ik\tau}}{2k^2\tau^2}.
\end{align}
It can easily be checked that they satisfy the normalization 
condition $|\alpha|^2-|\beta|^2=1$ at any time $\tau$. Their 
functional form is illustrated in Fig.~\ref{fig:bogoliubovcoefficientplot}.

Introducing $y:=-k\tau>0$, the Bogoliubov coefficients read
\begin{align}
    \alpha(y)= 1-\frac{1}{2y^2}+\frac{i}{y}, ~~~~ \beta(y) = \frac{e^{2iy}}{2y^2}.
\end{align}
We will also make use of the expressions
\begin{align}
|\alpha| &= \sqrt{1+\frac{1}{4y^4}}, \\
|\beta| & = \frac{1}{2y^2}, \\
\tan  {\rm arg}(\alpha) & =  \frac{2y}{2y^2-1} , \\
\tan  {\rm arg}(\alpha \beta ) & = \frac{2y \cos (2y)+(2y^2-1)\sin (2y)}{
(2y^2-1)\cos (2y) -2y \sin (2y)}.
\end{align}

Before analyzing the limiting cases, let us rewrite the expressions of the 
geometric complexity and gate complexity here:
\begin{widetext}
\begin{align}
    C[U_{\rm target}^{(1)}] &\lesssim 2\sqrt{{\rm arg}(\alpha(y))^2(1+4 
{\rm arsinh}^2|\beta(y)| 
   \csc^2(2 {\rm arg}(\alpha(y))))},\\ 
   D[U_{\rm target}^{(1)}] & =\frac{1}{\epsilon}\bigg[\bigg|2 {\rm arsinh}|\beta(y)|(\sin({\rm arg}(\alpha(y) \beta(y)))\bigg|+ \bigg|2 {\rm arsinh}|\beta(y)|\cos({\rm arg}(\alpha(y) \beta(y))))\bigg|+|2{\rm arg}(\alpha(y))|\bigg].
\end{align} 
\end{widetext}

It is instructive to study the UV ($|y| \gg 1$) and IR 
($|y| \ll 1$) approximations of the formulas. In the 
$y\rightarrow \infty$ limit, which corresponds to 
$\tau \rightarrow -\infty$, \emph{i.e.} early time, the Bogoliubov coefficients 
can be written as:
\begin{align}
    \alpha(y)=1, ~~~~ \beta(y)=0.
\end{align}
Because $\alpha$ and $\beta$ are both real, this regime is well within the
validity of our approximations. 
In this limit, the complexity of both approaches is given by 
\begin{align}
    C[U_{\rm target}^{(1)}] &\approx 0, \\
    \epsilon D[U_{\rm target}^{(1)}] & \approx 0\,.
\end{align}
This result is as expected, given the fact that 
in $y\rightarrow \infty$ limit, the de Sitter vacuum 
corresponds to the Minkowski vacuum, and hence, the 
target unitary operator is an identity. For the same reason
 modes that are well within the horizon have a negligible
 complexity of evolution.

 In the IR 
limit ($y \rightarrow 0$), the Bogoliubov 
coefficients can be written as
\begin{align}
    \alpha(y) = 1-\frac{1}{2y^2}+\frac{i}{y}, \\
    \beta(y) \approx -1+\frac{i}{y}+\frac{1}{2y^2}.
\end{align}
The inverse quadratic terms dominate, which are both real such that we are in
a regime of validity of our approximations corresponding to (\ref{Cr}).
Using $4 {\rm arsinh}^2|\beta| \csc^2(2 {\rm arg}(\alpha)) \gg 1 $,
Eq.~(\ref{complexityupperboundM}) can be simplified to
\begin{align}
\label{complexityIR}
    C[U_{\rm target}^{(1)}] & \lesssim  4 \mathcal{C}(y)|{\rm arsinh}|\beta||, \\
    \epsilon D[U_{\rm target}^{(1)}] & \approx |2 {\rm arsinh}|\beta|| +\mathcal{D}(y),
\end{align}
where the functions $\mathcal{C}(y)$ and $\mathcal{D}(y)$ are 
given by: 
\begin{align}
    \mathcal{C}(y) &= |{\rm arg}(\alpha) \csc(2{\rm arg}(\alpha))| = \frac{1}{2}+O(y^2), \\
    \mathcal{D}(y) &= |2 {\rm arg}(\alpha)| = 4|y|+O(y^3).
\end{align}
Furthermore, since ${\rm arsinh}|\beta| \approx -2 \ln |y|$, we find 
the following approximations for the upper bound for the geometric and 
gate complexities in the IR (super-Hubble) limit:
\begin{align}
\label{complexityIRFInal}
    C[U_{\rm target}^{(1)}] & \approx 4 |\ln |y|| \sim \ln a \sim t, \\
    \epsilon D[U_{\rm target}^{(1)}] & \approx 4 |\ln |y|| \sim \ln a \sim t,
\end{align}
where the relation $a = - 1/(H\tau) = a_0e^{Ht}$ has been used. 
Therefore, both types of complexity consistently predict a 
logarithmic increase of complexity  as a function of  the scale factor $a$
for the super-Hubble domain.   

The rate of change of complexity in coordinate time $t$, 
in the IR limit is given by:
\begin{align}
    \frac{dC[U_{\rm target}^{(1)}]}{dt} & \sim \frac{\dot{a}}{a} = 
    \sqrt{\frac{\Lambda}{3}} = \text{const}, \\
    \frac{d(\epsilon D[U_{\rm target}^{(1)}])}{dt}  & \sim \frac{\dot{a}}{a} = \sqrt{\frac{\Lambda}{3}} = \text{const}.
\end{align}
The complexity change rate (with respect to the coordinate time) is thus 
constant and proportional to the Hubble expansion rate, given by 
$H=\sqrt{\frac{\Lambda}{3}}$. This observation provides a possible 
interpretation of the cosmological constant as a measure of computational 
performance. The computational performance (given by the time derivative 
of complexity) is needed to achieve a state of complexity predicted 
for the de Sitter model.

\begin{figure}[h!]
    \centering
    \includegraphics[scale=0.5]{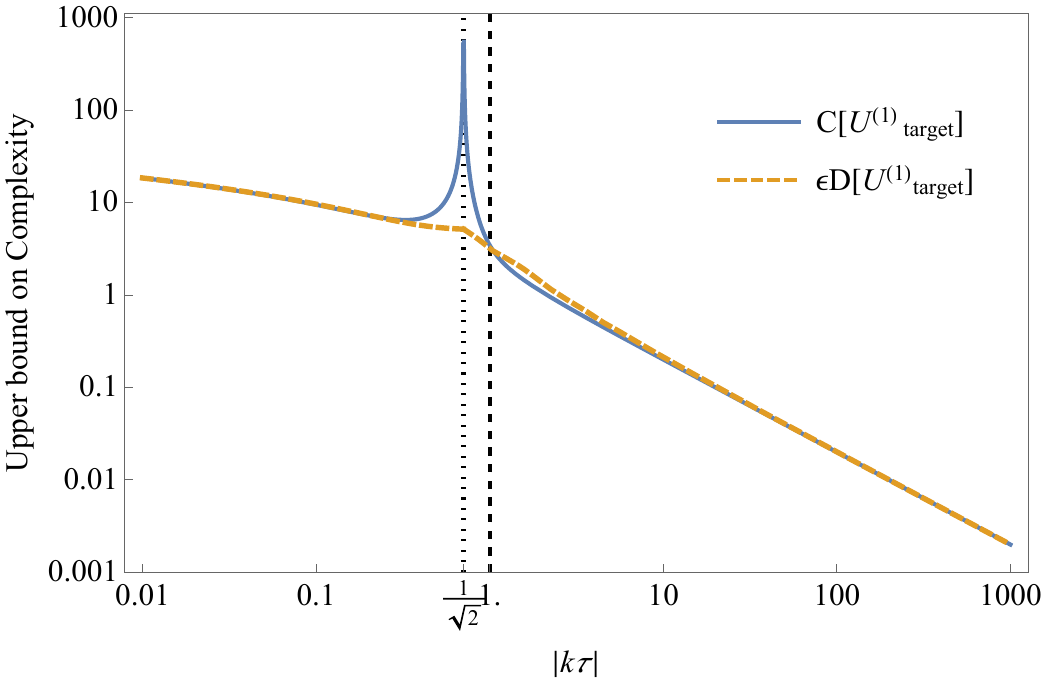}
    \caption{Log-log plot of the upper bounds on the complexity (computed geometrically 
    and via a gate counting approach) of a scalar field mode in de Sitter spacetime. 
    The vertical dashed line separates the super-Hubble (left) and sub-Hubble (right) 
    regions. The vertical dotted line indicated the location of the peak at $|k\tau|=\frac{1}{\sqrt{2}}$.}
    \label{fig:complexitydesitter}
\end{figure}

The full behavior of the upper bound on complexity and the gate complexity 
(up to a scaling factor $\epsilon$ in the latter case) is shown in Fig.~\ref{fig:complexitydesitter}. 
At early times $|k\tau| \gg 1$ (sub-Hubble modes), 
 the value of complexity 
is small. For super-Hubble modes, the complexity exhibits 
a growing behavior. In the transition region of $|k\tau|\approx 1$, there are
significant differences between the geometrical and gate complexities, but in
this regime, our approximations for the geometrical complexity are not
reliable. They may still be used as an upper bound, but it is not necessarily
close to the correct result. Indeed, the gate complexity is much smaller than
this upper bound in a large part of the transition region.  However, there is
also a small range of $|k\tau|>1$ in which the gate complexity is larger than
our upper bound obtained from geometrical complexity. This is an indication
that the gates used in our derivation of the gate complexity can be improved.

Formally, an inspection of our equations shows that the peak of the
upper bound on geometrical complexity is due to the term $\csc^2(2 {\rm arg}(\alpha))$
in
\begin{align}
C[U_{\rm target}^{(1)}] \lesssim 2\sqrt{{\rm arg}^2(\alpha)(1+4 
{\rm arsinh}^2|\beta| 
   \csc^2(2 {\rm arg}(\alpha)))}.
\end{align}
As a function of $y$, this expression is:
\begin{equation}
\csc^2(2 {\rm arg}(\alpha)) = \frac{\left(4 y^4+1\right)^2}{16 y^2 \left(1-2 y^2\right)^2}.
\end{equation}
The peak is associated with the divergence of the expression at: 
\begin{equation}
|k\tau|= |y| = \frac{1}{\sqrt{2}} \approx 0.707. 
\end{equation}

\begin{figure}[h!]
    \centering
    \includegraphics[scale=0.5]{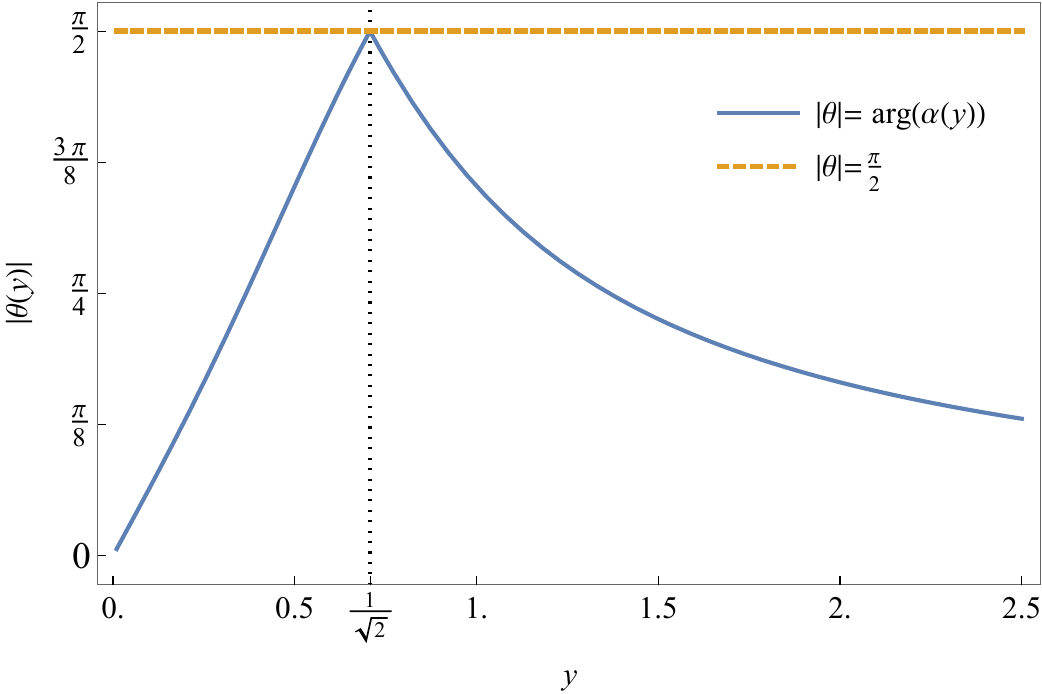}
    \caption{Variation of $|\theta|$={\rm arg}($\alpha$) as a function of $y$. 
    The vertical dotted line at $y=\frac{1}{\sqrt{2}}$ indicated the location 
    of the complexity peak, which corresponds to $|\theta|=\frac{\pi}{2}$.}
    \label{thetaplot}
\end{figure}

The value $|y|=\frac{1}{\sqrt{2}}$ corresponds to the point where 
$|\theta|= \pm \frac{\pi}{2}$ (as can be deduced from Fig. \ref{thetaplot}), which 
is where equations (\ref{eqn1v}) and (\ref{eqn2v}) do not hold. This indicates that the 
divergence of complexity at $|y|=\frac{1}{\sqrt{2}}$ is merely a mathematical artifact 
stemming from the upper bound approximation used in our analysis.

The characteristic difference in the behavior of geometrical complexity a
cross the horizon indicates that it could be used as a probe to capture information about 
the underlying cosmological spacetime and its horizon.
Geometrical complexity may, therefore, encode information about
the structure of the underlying spacetime. However, this transition regime is
sensitive to higher-order corrections in both the Dyson series and the BCH
formula.  The qualitative behavior of $C[U_{\rm target}^{(1)}]$ is preserved
if we add a higher-order contribution to the BCH expansion, as shown in
Appendix~\ref{appC}.  It is also important to note that the existence of the
peak of the approximate upper bound does not imply a divergence of the
complexity itself. For now, it is simply more difficult to resolve the precise
complexity in this regime.  Moreover, if there is room for the complexity to
increase at this point, physical arguments such as Lloyd's bound
\cite{Lloydbound} may suggest that this freedom should not be utilized to its
full potential.

In \cite{Bhattacharyya:2020rpy, Bhattacharyya:2020kgu}, the complexity for a
cosmological perturbation in a dS cosmological background, described in terms
of the two-mode squeezed state, was explored relative to the two-mode vacuum
state. The authors found that the complexity remains small while the mode is
within the Hubble radius and linearly increases with the log of the scale
factor after the horizon exit. A similar observation of the complexity growth
of mode functions of inflationary perturbations was made in
\cite{Lehners:2020pem}. These results are in qualitative agreement with our
findings. However, neither of these previous studies noticed the subtle
transition behavior around the peak we observed here, nor has an analysis of
the gate complexity been made.

\section{Discussions and Summary}
\label{sec7}

The complexity of quantum processes in itself is an interesting question 
and is worth investigating. In the present work, we set out to determine 
the complexity of the evolution of scalar field modes in de Sitter spacetime.
We adopted the geometrical approach suggested by Nielsen \textit{et al.\ }in
their pioneering works in which the complexity of an operation can be  
related to the geodesic distance between the identity operator and 
the desired unitary in a suitable unitary group manifold. 

As a cosmological application, the behavior of a scalar field mode in a time-dependent background can be 
explained by considering an oscillator with a time-dependent 
frequency. We first analyzed this underlying model in general terms. In order to
apply the geometric approach by Nielsen, we expressed the unitary 
operator associated with the evolution of a time-dependent harmonic 
oscillator as a product of two unitary operators, which are popularly 
known as the squeezing and the rotation operators. Having expressed the 
time evolution operator in this way, we identified the set of fundamental
operators that can be used to construct
the desired target unitary operator. The fundamental operators correspond 
to a certain Lie group, which determines the geometry in which we should look 
for geodesics to compute the complexity. The geodesics in the Lie group 
manifold were obtained by solving the Euler--Arnold equation, placing equal
penalties on the group generators.

The Euler--Arnold equation directly determines the tangent vector field along
a geodesic.
The actual trajectory on the group manifold, starting at some initial point
and continued in the direction of the tangent vector, can be expressed as
a path-ordered exponential. Explicitly computing this object is a cumbersome
task, usually simplified by approximating the required expression 
following an iterative approach. The result is a Dyson series, which we
restricted in our computations to the leading-order term. Since we are not
computing the exact geodesic, the length along our solution curve
places an upper bound on the complexity instead of providing
a precise value. We also included a second approximation based on the
Baker--Campbell--Hausdorff formula. This approximation slightly changes the
target unitary and may, therefore, result in a smaller or larger distance,
depending on the specific case. The combination of both expansions provides an
approximate upper bound on the complexity. The approximation can be viewed as
an expansion in $\hbar$, or a loop expansion because higher-order terms are
determined by iterations of commutators.

We showed that the upper bound on complexity can be written explicitly in
terms of the Bogoliubov coefficients of transformations between ground states
in two different regimes. In order to test our results for the geometrical
complexity, we also determined the gate complexity, an approach that involves
counting the minimum number of gates required to construct a given
operator. Our first example was the model of an oscillator with switched
frequency where the time-dependent frequency profile exhibited an
instantaneous change from $\omega_{\rm in}$ to $\omega_{\rm out}$. We studied
how the complexity behaves as a function of
$\omega_{\rm in}/\omega_{\rm out}$. When
$\omega_{\rm in}=\omega_{\rm out}$, the complexity vanishes, which is as
expected as the reference and the target state are identical in this limit.
Hence, the target unitary operator is an identity whose complexity does vanish.

We then considered a frequency profile that has direct relevance in the
context of cosmology, describing the mode of a free quantum field in de Sitter
spacetime. In this case, our motivation was to find the complexity of the time
evolution of a field mode and possible relationships with the de Sitter
horizon. We observed that the geometrical and gate complexities lead to
equivalent results in the UV and IR regimes. The complexity computed by the two
different approaches is negligible when the mode is within the Hubble radius
(sub-horizon modes) and grows as the logarithm of the scale factor for
super-horizon modes. However, the geometrical complexity seems much more
sensitive during the horizon transition, exhibiting a sharp peak near the 
Hubble horizon. Notably, the feature was observed for the leading-order
contribution to the Dyson series employed in this article. The robustness 
of these predictions beyond the leading-order terms requires further 
investigation, which will be addressed in future studies.

This analysis opens new avenues for exploring quantum 
complexity in cosmological systems and beyond. Such investigations hold 
significant importance for multiple reasons. Firstly, they could reveal 
novel computation-related aspects of cosmological evolution, such as whether 
the evolution of quantum inhomogeneities in cosmological backgrounds achieves 
computational optimality. Additionally, they provide an opportunity to test 
hypotheses regarding the growth of complexity in the Universe. Secondly, 
analyzing complexity offers a practical upper bound for estimating the 
computational resources required to simulate the quantum processes under 
study. Expanding the scope of this research to other cosmological scenarios 
could further establish the utility of geometrical quantum complexity as a 
powerful framework for characterizing the fundamental properties of 
cosmological evolution.

\begin{acknowledgments}
SC would like to thank the Doctoral School of Exact and Natural Sciences of 
Jagiellonian University for providing fellowship during the course of the work. 
The research was conducted within the Quantum Cosmos Lab (\href{https://quantumcosmos.org}
{https://quantumcosmos.org}) at the Jagiellonian University. The work of MB
was supported in part by NSF grant PHY-2206591.

\end{acknowledgments}

\appendix

\section{Expressing the time evolution operator as a product of the squeezed 
and the rotation operator.}
\label{appB}
In this appendix, we will derive the time evolution operator as a product of 
two unitary operators. The creation and the annihilation operator at some time 
$t$ can be expressed in terms of the operators at initial time using the 
Bogoliubov coefficients, which, under a suitable parametrization, can be written 
as 
\begin{align}
\label{bogoliubivsqueezedapp}
    {a}(t) &= e^{i \theta(t)}\cosh(r(t)) {a}_0- e^{i(\phi(t)-\theta(t)}\sinh(r(t)) {a}^{\dagger}_0, \\
    {a}^{\dagger}(t) &= e^{-i \theta(t)}\cosh(r(t)) {a}_0- e^{-i(\phi(t)-\theta(t))}\sinh(r(t)) {a}^{\dagger}_0. 
\end{align}

This Bogoliubov transformation can be represented as a unitary transformation,
\begin{align}
    {a}(t)= U^{\dagger}(t){a}(t_0) U(t).
\end{align}
with a unitary operator $U(t)$. Our aim is to show that $U(t)$ can be
expressed as the product of two operators,
\begin{align}
    U(t)= S(r(t),\phi(t))R(\theta(t)),
\end{align}
with separate dependencies on the parameters.
To do so, we begin with the definition of the unitary operator $S(r,\phi)$ by
\begin{equation}
    S = \exp\bigg(\frac{1}{2}r e^{-i \phi} a^2-\frac{1}{2}r e^{i \phi}
    a^{\dagger 2}\bigg)
  \end{equation}
  such that
  \begin{equation}
    S^{\dagger} = \exp\bigg(\frac{1}{2}r e^{i \phi} a^{\dagger 2}-\frac{1}{2}r e^{-i \phi} a^{2}\bigg).
\end{equation}
Here, $r(t)$ and $\phi(t)$ are simply written as $r$ and $\phi$ for the sake
of notational simplicity, and we will follow this convention hereafter.

Denoting
\begin{equation}
  \frac{1}{2}r e^{-i \phi} a^2-\frac{1}{2}r e^{i \phi} a^{\dagger 2}= J\,,
\end{equation}
we  write  
\begin{align}
    S= \exp(J) ~~~~ {\rm and}~~~  S^{\dagger}= \exp(-J). 
\end{align}

This implies,
\begin{align}
\label{eqn1}
    S^{\dagger} a S= \exp(-J) a \exp(J). 
\end{align}
Applying the Baker-Campbell-Hausdorff (BCH) formula
\begin{align}
    e^{\lambda B} A e^{-\lambda B}= A+ \lambda [B,A]+ \frac{\lambda^2}{2!}[B,[B,A]]+\cdots
\end{align}
we can write equation (\ref{eqn1}) as 
\begin{align}
    S^{\dagger} a S= a - [J, a]+ \frac{1}{2!}[J,[J,a]]-\cdots
\end{align}
Furthermore, it can be checked that
\begin{align}
    [J, a]= r e^{i \phi} a^{\dagger}.
\end{align}
Therefore,
\begin{align}
    S^{\dagger} a S & = a- r e^{i \phi} a^{\dagger}+ \frac{1}{2!}r^2 a- \frac{1}{3!} r^3 e^{i \phi} a^{\dagger}+ \frac{1}{4!}r^4 a - \cdots \\
    & = a(1+ \frac{r^2}{2!}+ \frac{r^4}{4!}+\cdots)- a^{\dagger}e^{i \phi}(r+ \frac{r^3}{3!}+\cdots) \\
    &= a \cosh(r)- e^{i \phi} a^{\dagger} \sinh(r).
\end{align}
Now, 
\begin{align}
    R^{\dagger}S^{\dagger} a S R &= R^{\dagger} \bigg(a \cosh(r)- e^{i \phi} a^{\dagger} \sinh(r)\bigg) R \\
    & = \cosh(r) R^{\dagger} a R - e^{i \phi} \sinh(r) R^{\dagger} a^{\dagger}R.
\end{align}
Again, using the BCH formula, we can prove 
\begin{align}
    R^{\dagger} a R &= e^{i\theta} a, \\
    R^{\dagger} a^{\dagger}R &= e^{-i \theta} a^{\dagger},
\end{align}
which gives:
\begin{align}
\nonumber
    R^{\dagger}S^{\dagger} a S R &= \cosh(r) e^{i \theta} a - e^{i \phi} \sinh(r) e^{-i \theta}a^{\dagger} \\
    &= e^{i \theta} \cosh(r) a - e^{i(\phi-\theta)}\sinh(r) a^{\dagger}.
\end{align}

Following the previous procedure, the transformation equation for $a^{\dagger}$ can be derived as follows:
\begin{align}
\nonumber
    S^{\dagger} a^{\dagger}S &= a^{\dagger}-[J,a^{\dagger}]+\frac{1}{2!}[J,[J,a^{\dagger}]]-\cdots \\ \nonumber
    &= a^{\dagger}- r e^{-i \phi} a+ \frac{1}{2!}r^2
      a^{\dagger}-\frac{1}{3!}r^3 e^{-i\phi}a + \frac{1}{4!}r^4 a-\cdots  \\
    &= \cosh(r)a^{\dagger}- \sinh(r) e^{-i\phi}a.
\end{align}
Therefore,
\begin{align}
    R^{\dagger}S^{\dagger} a^{\dagger} S R= e^{-i \theta}\cosh(r) {a}_0- e^{-i(\phi-\theta)}\sinh(r) {a}^{\dagger}_0.
\end{align}

\section{An improvement on the upper bound of complexity for the out target unitary}
\label{appC}

In this appendix, we derive the contribution of the nested commutator term in
the operator expansion of the complexity, based on the BCH formula. As
described in Section~\ref{subsec:Nielsen}, our original target unitary is of the
form
\begin{align}
    U_{\rm target} = e^{X}e^Y,
\end{align}
with, in general, non-commuting $X$ and $Y$.  In order to implement the
boundary condition, it is necessary to express the product of exponentials as
a single exponential, which can be done by using the BCH formula. The BCH
formula includes the contributions of all the nested commutators. In the main
text, we neglected the nested commutators and computed the complexity of
\begin{align}
    U_{\rm target}^{(1)} \approx e^{X+Y}.
\end{align}
We now include the contribution of the commutator term in the BCH expansion,
leading to a new approximation of the target operator that we will denote as
$U_{\rm target}^{(2)}$. We will then determine the difference in the
complexity upper bound compared with
the previously obtained result.

Including the first commutator term according to the BCH formula, the target
operator can be written as
\begin{align}
    U_{\rm target}^{(2)} \approx e^{X+Y+\frac{1}{2}[X,Y]}\,.
\end{align}
Implications of this new term are easy to derive, and one could use similar
methods to proceed to even higher orders. However, a complete higher-order
treatment would have to include higher terms in the Dyson series because their
magnitude is also determined by the commutator structure of the Lie
algebra. These terms are more challenging and will be addressed elsewhere.

For our $X=-2i r(t)(\sin(\phi(t))\mathcal{O}_1+\cos(\phi(t))\mathcal{O}_2$ and $Y=2i\theta(t)\mathcal{O}_3$
used earlier in (\ref{originaltarget}), this operator equals
\begin{widetext}
\begin{align}
\nonumber
    U_{\rm target}^{(2)} & = \exp\bigg(-2i r(\sin \phi \mathcal{O}_1+\cos \phi \mathcal{O}_2)+2i \theta \mathcal{O}_3-2 i r \theta (\sin \phi \mathcal{O}_2-\cos \phi \mathcal{O}_1)\bigg) \\
    & = \exp\bigg\{-2ri( \sin \phi- \theta \cos \phi)\mathcal{O}_1-2ri(
      \cos \phi +   \theta \sin \phi)\mathcal{O}_2 + 2i \theta \mathcal{O}_3
      \bigg\} \,.
\end{align}
\end{widetext}
New terms implied by the commutator can easily be identified by the products
of $\theta$ times trigonometric functions of $\phi$.

The values of the $v_I$ are determined by $U(s=1)=U_{\rm
  target}^{(2)}$. Comparing the coefficients of the generators $\mathcal{O}_I$
implies: 
\begin{align}
    v_3 &=-2 \theta,  \\
    \frac{v_1 \sin(2 v_3)}{2 v_3}-\frac{v_2 \sin^2(v_3)}{v_3} &= 2 r \sin \phi-2 r \theta \cos \phi,  \\
    \frac{v_1 \sin^2(v_3)}{v_3}+\frac{v_2 \sin(2 v_3)}{2 v_3} &= 2 r \cos \phi 
    + 2 r \theta \sin \phi. 
\end{align}
and therefore
\begin{align}
   v_3 &= -2\theta,\\ 
   v_1 &= -4 \theta  r \csc (2 \theta) (\sin (2 \theta -\phi)+\theta  \cos (2 \theta -\phi)), \\
   v_2 &= -4 \theta  r \csc (2 \theta) (\theta  \sin (2 \theta -\phi)-\cos (2 \theta-\phi)).
\end{align}
The upper bound on the complexity is now given by:
\begin{align}
    C[U_{\rm target}^{(2)}] \lesssim 2\sqrt{\theta(t)^2(1+4r(t)^2(1+\theta(t)^2)\csc^2(2\theta(t)))}\,.
\end{align}

We can already see that the correction, contained solely in the factor of
$1+\theta(t)^2$, is subdominant in the previous range of validity where
$\theta(t)$ was required to be small. The correction is relevant in
any regime where $\theta$ is of the order one or larger. However, it always
increases the result and, therefore, does not sharpen our leading-order
upper bound.

\begin{figure}[ht!]
    \centering
    \includegraphics[scale=0.5]{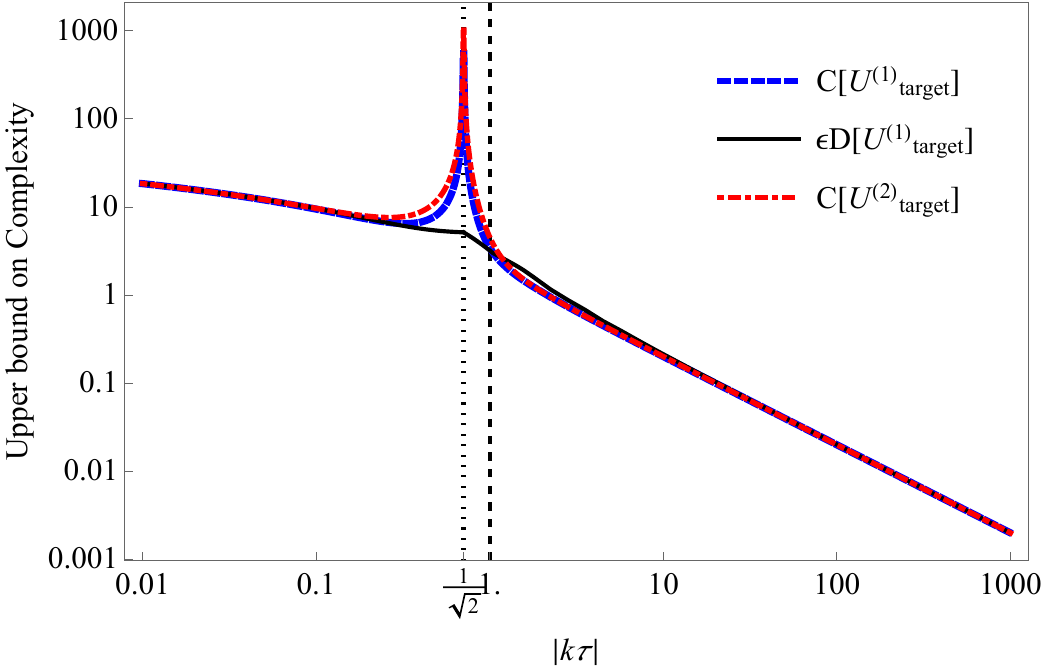}
    \caption{Quantum complexities under consideration, as functions of
      $|k\tau|$. The first commutator correction slightly increases our
      leading-order result in the transition region where $|k\tau|$ is of order one.}
    \label{fig:desittercomplexityimproved}
\end{figure}

In terms of the Bogoliubov coefficients, the new upper bound can be written as: 
\begin{widetext}
\begin{align}
    C[U_{\rm target}^{(2)}] &\lesssim \frac{1}{\sqrt{2}}|\arg(\alpha) \csc(\arg (\alpha))\sec(\arg(\alpha))|\nonumber \\ 
   &\times \sqrt{(1-\cos(4 \arg(\alpha))+8 {\rm
     arsinh}^2|\beta|(1+\arg^2(\alpha)))} \nonumber \\
  &=2|\arg(\alpha)| \sqrt{1+4{\rm
     arsinh}^2|\beta|(1+\arg^2(\alpha))\csc^2(2\arg (\alpha))}\,.
\end{align}
\end{widetext}
The UV and the IR limits can be derived as in the main text, without
significant differences because they correspond to small or vanishing
$\theta$.  Figure~\ref{fig:desittercomplexityimproved} illustrates a
comparative analysis of the behavior of gate complexity, the geometrical upper
bound, and the new version obtained here.  

 The new version of the upper bound
exhibits the same peak as the previous upper bound, which is not captured by
the gate complexity. However, the behavior in the UV and the IR limits remains
similar. We note that the upper bound and the new version do not show any
significant differences in their behavior, suggesting that the addition of the
commutator term in the operator expansion does not affect the complexity in a
significant way. In fact, as already expected from the analytical expression,
the correction increases the previous upper bound and, therefore, does not
sharpen it. However, the contribution of additional nested commutator terms might
result in observable changes in the complexity behavior, and, probably more
importantly, higher orders in the Dyson series would also have to be included.

\bibliography{biblio}
\bibliographystyle{utphys}


\end{document}